\documentclass[12pt,a4paper]{article}
\usepackage{epsf,epsfig,amsmath,colordvi,graphicx,color,xr}
\begin{document}
\vspace{20mm}
\setlength{\baselineskip}{14pt}
\begin{center}
{\Large\bf\uppercase{Velocity, Velocity Gradient and Vorticity Statistics of Grid Turbulence obtained using Digital Cross-Correlation PIV}}\\
\vspace{10mm}
{\normalsize By PHILIPPA O'NEILL, DAVID NICOLAIDES, DAMON HONNERY \small{AND} \normalsize JULIO SORIA}\\
\vspace{5mm}
\setlength{\lineskip}{14pt}
{\normalsize Laboratory for Turbulence Research in Aerospace \& Combustion\\ Department of Mechanical and Aerospace Engineering Monash University\\ Melbourne, Clayton, VIC 3800, Australia}\\
Contact: julio.soria@monash.edu
\end{center}

\begin{abstract}
\setlength{\lineskip}{10pt}
\noindent 
Grid turbulence is investigated using cross-correlation digital Particle Image Velocimetry (PIV) over a range of Taylor Reynolds Number ($Re_\lambda$) from 5 to 44. Instantaneous velocity is measured directly and vorticity and velocity gradients are obtained indirectly. Measurements are taken at various downstream locations from the generating grid. Probability distribution functions (PDFs) are calculated for the fluctuating component of the velocity, the spatial velocity gradients and vorticity. The PDF of the velocity fluctuations has a Gaussian distribution while velocity gradients and vorticity are found to have non-Gaussian PDF distributions. 

The structure of the flow is investigated by calculating spatial autocorrelations for all measured and derived data. The spatial velocity autocorrelations differ from previous experimental measurements of grid turbulence, most of which have been determined from single-point measurements. This difference is believed to be due to differences in the way in which the measurements are made, and particularly to differences in the spatial size of the experimental domain.
 
\end{abstract}

\section{Introduction}
A statistical description of turbulent motion is greatly simplified if the flow field is homogeneous and isotropic.  Grid turbulence is an approximation to homogeneous, isotropic turbulence and has been widely studied (eg. Batchelor and Townsend (1947), Champagne et.al. (1970), Frenkiel et.al. (1979), Kit et.al. (1988), Tsinober et.al. (1992)). It is generated when a grid constructed from overlaid orthogonal cylinders is placed in the flow normal to the flow direction. There is a convective velocity, and therefore grid turbulence is not strictly homogeneous and isotropic. Immediately following the grid there is a developing region where the flow is inhomogeneous and anisotropic (Mohamed and LaRue (1990)). This region is followed by one where the flow is nearly homogeneous and isotropic. This region is know as the 'initial period of decay' and was defined by Batchelor and Townsend (1948) as being within the range $20\le x/M \le 100$, where $x$ is the distance downstream from the grid and $M$ is the spacing between grid elements. Generally, flow measurements have been taken in this region using equipment such as hot-wires. Temporal single point measurements are obtained from which the temporal velocity gradient is determined, and this can then be transformed into a spatial velocity gradient using Taylor's hypothesis (Townsend (1947)). However, the application of Taylor's hypothesis introduces some uncertainty into the measurement (Comte-Bellot and Corrsin (1971), Antonia et.al. (1980)). The direct method for determining spatial velocity gradients is to calculate them from spatial velocity measurements, in which case a technique is needed that is capable of obtaining closely spaced spatial measurements. Such a technique is PIV.

PIV is a flow measurement technique that provides, in its most basic form, two components of velocity over a two-dimensional domain (2C-2D PIV) (Willert and Gharib, 1991). Spatial velocity gradients can then be calculated and, as out-of-plane vorticity is a function of in-plane spatial velocity gradients, out-of-plane vorticity can also be calculated. In the case of grid turbulence, the three dimensional velocity field is fully characterised by measuring the convective stream-wise component and one of the cross-stream components.

In this paper, statistics of the velocity gradients and out-of-plane vorticity in grid turbulence are reported, along with velocity statistics. Experimentally determined velocity, velocity gradient and vorticity statistics are also compared to numerical results.

\section{Experimental Apparatus and Methodology}
\subsection{Experimental Apparatus}
Experiments are carried out in a vertical closed-circuit water tunnel. The water in the tunnel is driven by a 3kW stainless steel centrifugal pump, controlled via a frequency inverter. This allows the flow speed in the working section of the tunnel to be maintained. The internal dimensions of the working section are 250 mm x 250 mm; it is 1500 mm long and constructed from 15 mm thick acrylic sheet. The settling chamber is located above the working section and consists of four turbulence damping screens made of stainless steel and one honeycomb section made of plastic. Water enters the settling chamber through a spray system, and the water in the settling chamber enters the working section via a 16:1 contraction. The free-stream natural turbulence level is generally less than 1\% for all the experimental conditions investigated, and tends to be higher further downstream from the contraction and at higher free-stream velocities. Further details can be found in Nicolaides (1997).

The particles for PIV are 11 $\mu$m hollow glass spheres and have a specific gravity of 1.1. Illumination of the particles in the image plane is provided by two Spectra Physics 400 mJ pulsed Nd:YAG lasers with an optimum firing frequency of 12 Hz. The laser beam is spread into a laser sheet by a series of optics. The camera is a Kodak Megaplus XHF camera with 1 million pixels on the CCD array (1000 (V) x 1000 (H)) and 8 bit resolution. The maximum framing rate for the system is 22 Hz. Timing of the laser and camera is controlled by a program written in-house. Both single and double-exposed images were acquired for PIV analysis. Single exposed images were preferred, but at higher water tunnel speeds results could only be obtained using double exposed images. The cross-correlation PIV algorithm used to analyse the image pairs is described in Soria (1996a, b). It uses an adaptive technique to increase the velocity dynamic range and reduce the bias and random errors in comparison to standard cross-correlation PIV analysis.

Velocity measurments are made over an area approximately 3$\lambda$ x 3$\lambda$, with a spatial resolution of between 0.07$\lambda$ and 0.1$\lambda$. The Taylor microscale $\lambda$, is estimated from:
\begin{equation}
\lambda = \sqrt{\frac{{\bf{u}}^2}{\langle{(\frac{{\partial u}}{\partial x}})^2\rangle}}
\label{taylor-1}
\end{equation} 
where $\bf{u}$ is the root mean square of fluctuating velocity and $u$ is the fluctuating component of velocity in the $x$ or stream-wise direction. The spatial resolution in this study compares favourably to the multi hot-wire measurements in grid turbulence of Tsinober et al. (1992), where the spatial resolution obtained was between 0.5$\lambda$ and 0.67$\lambda$. 

\subsection{Experimental Conditions}
Two different grids are used in the experiments, and their properties are presented in Table \ref{tab:gridprops}. All measurements are taken along the central plane of the working section, at a number of locations downstream from the grid. These locations are denoted by symbols $x1$, $x2$ etc, and are given in Table \ref{tab:exploc}. The spatial resolution for all measurements acquired using grid A is 30 $\mu$m/pixel, or 33.33 pixels/mm, and for grid B is 25 $\mu$m/pixel, or 40 pixels/mm.

\begin{table}
\begin{center}
\begin{tabular}{|c|c|c|c|c|}\hline
Grid & \it{M} (mm) & \it{d} mm & $M/d$ & Start of initial period of decay:$x/M \approx 10$ \\ \hline
A & 30 & 6 & 5 & $\approx 300$ mm\\
B & 15 & 3 & 5 & $\approx 150$ mm \\
\hline
\end{tabular} 
\caption{Properties of turbulence generating grids}
\label{tab:gridprops}
\end{center}
\end{table}

The timing of image acquisition for PIV analysis falls into one of three categories, listed in Table \ref{tab:acq}. In this table $\Delta t_1$ is the time delay between the image and its preceeding image and $\Delta t_2$ is the time delay beween the image and its succeeding image. Table \ref{tab:expcond} summarises the grid conditions and image acquisition parameters for the experiments, while the flow conditions are summarised in Table \ref{tab:flowcond}.

\begin{table}
\begin{center}
\begin{tabular}{|c|c|c|c|}\hline
Location & Distance downstream from grid (mm) & $x/M$ ($M=30$ mm) & $x/M$ ($M=15$ mm)  \\ \hline
x1 & 500 & 17 & 33 \\
x2 & 600 & 20 & 40 \\
x3 & 700 & 23 & 47 \\
x4 & 900 & 30 & 60 \\
x5 & 1000 & 33 & 67 \\
x6 & 1050 & 35 & 70 \\
\hline
\end{tabular} 
\caption{Downstream location of imaged areas, with distance from the grid in terms of spacing of the grid elements}
\label{tab:exploc}
\end{center}
\end{table}

\subsection{Velocity Error}
Errors in the separation time between correlated images and errors in determining the dimensions of the imaged plane are random errors that will effect the determination of velocity. The resolution of the time separation between images is 0.001 ms, and the time separation is greater than 1 ms for all the experiments performed. The error due to the time separation is therefore, at most $\pm$ 0.1\%. The error in the measurement of each dimension of the imaged plane is estimated to be less than 0.25 mm, which is equivalent to $\pm$ 1.0\% for data with a spatial resolution of 40 pixels/mm. The combined random error on the measured velocity is therefore, at most, $\pm$ 1.1\%.

\begin{table}
\begin{center}
\begin{tabular}{|c|c|c|}\hline
Acquisition type & $\Delta t_1$ (ms) & $\Delta t_2$   \\ \hline
fast & 34.0 & 90.2 \\
slow & 45.2 & 45.2 \\
double-exposure & 83.3 & 83.3 \\
\hline
\end{tabular} 
\caption{Timing for the various image acquisition modes. $\Delta t_1$ is the time delay between the image and its preceeding image,  $\Delta t_2$ is the time delay between the image and its succeeding image}
\label{tab:acq}
\end{center}
\end{table}

The predominant error in the velocity measurements is the random error in determining the location of the peak in the cross-correlation function for PIV. For the PIV technique employed in these experiments, this error has been shown to be random and have an approximate Gaussian distribution (Soria (1996a, b)). This error is denoted as $\sigma_{piv}$, and is the standard deviation of the random Gaussian error.

\begin{table}
\begin{center}
\begin{tabular}{|c|c|c|c|c|c|c|}\hline
Data & Grid & No. of images & Acquisition & Acquisition  & No. of & No. of vectors \\
set &  & acquired & type & location  & vector fields & per field  \\ \hline
A1 & A & 63 & slow & x2 & 62 & 46 x 59 = 2714 \\
A2 & A & 63 & slow & x3 & 62 & 46 x 59 = 2714 \\
A3 & A & 59 & fast & x4 & 28 & 46 x 59 = 2714 \\
A4 & A & 63 & fast & x5 & 31 & 46 x 59 = 2714 \\ \hline
B1 & A & 63 & double & x1 & 62 & 26 x 28 = 728 \\
B2 & A & 63 & double & x1 & 62 & 26 x 28 = 728 \\
B3 & A & 63 & double & x1 & 62 & 26 x 28 = 728 \\
B4 & A & 63 & double & x1 & 62 & 26 x 28 = 728 \\ \hline
C1 & B & 200 & fast & x1 & 99 & 48 x 59 = 2832 \\
C2 & B & 200 & fast & x2 & 99 & 48 x 59 = 2832 \\
C3 & B & 200 & fast & x6 & 99 & 48 x 59 = 2832 \\ \hline
D1 & B & 100 & double & x1 & 99 & 24 x 28 = 672 \\
D2 & B & 100 & double & x4 & 99 & 24 x 28 = 672 \\
D3 & B & 100 & double & x6 & 99 & 24 x 28 = 672 \\ \hline
E1 & B & 100 & double & x1 & 99 & 24 x 28 = 672 \\
E2 & B & 100 & double & x2 & 99 & 24 x 28 = 672 \\
E3 & B & 100 & double & x4 & 99 & 24 x 28 = 672 \\
E4 & B & 100 & double & x6 & 99 & 24 x 28 = 672 \\ \hline
F1 & B & 100 & double & x1 & 99 & 24 x 28 = 672 \\
F2 & B & 100 & double & x2 & 99 & 24 x 28 = 672 \\
F3 & B & 100 & double & x4 & 99 & 24 x 28 = 672 \\ \hline
\end{tabular} 
\caption{Experimental and images acquisition parameters for each data set}
\label{tab:expcond}
\end{center}
\end{table}

\begin{figure}[t!]
\centering
\includegraphics[width = 120mm, angle = 0]{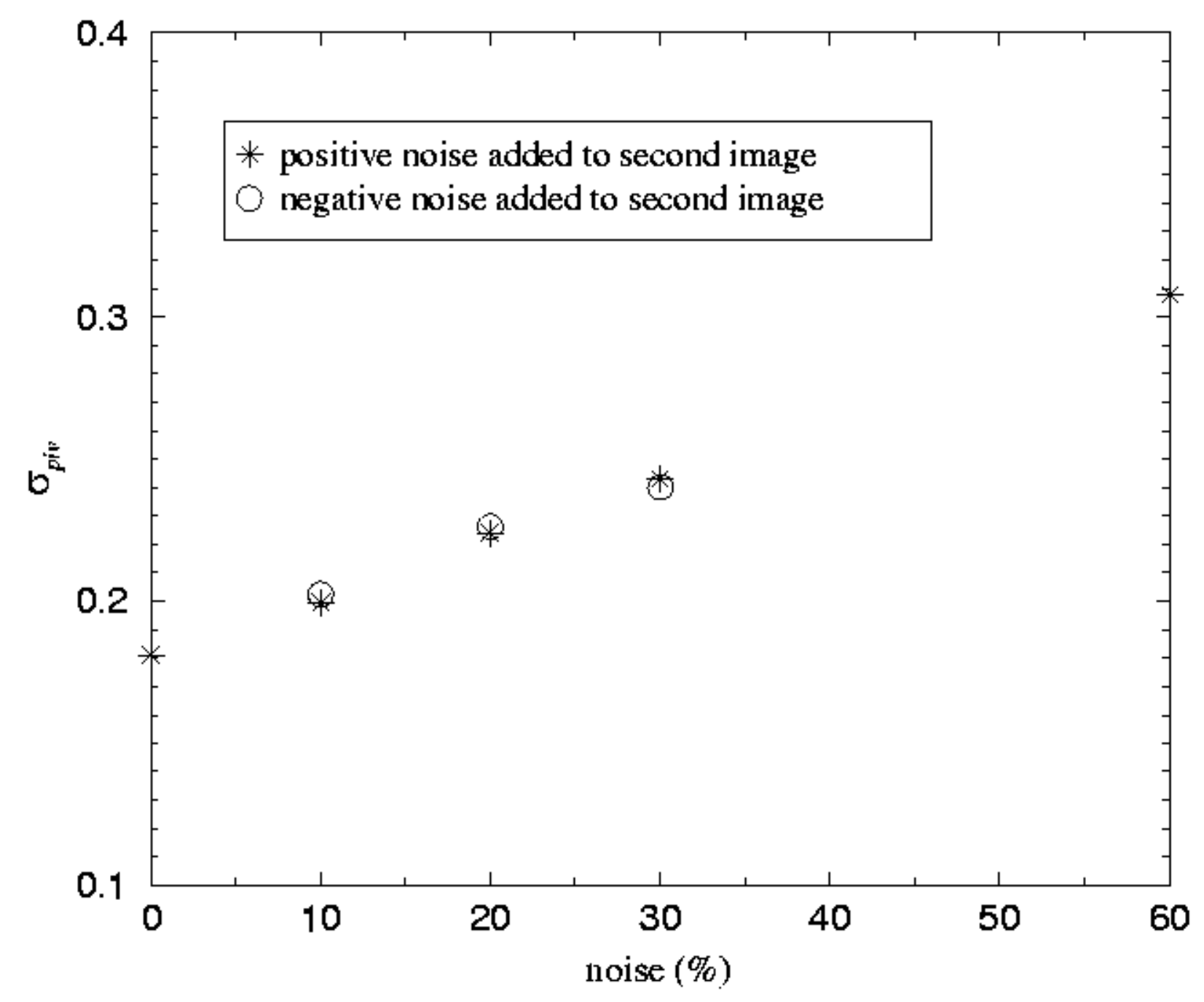}\\
\caption{Effect of uniformly distributed noise on the standard deviation for a 32 pixel x 32 pixel analysis region. The percentage noise added is with respect to 256, which is the dynamic range of the camera.}
\label{fig:error-piv}
\end{figure}

In order to determine the sensitivity of the cross-correlation analysis to images with background noise, uniformly distributed noise was added randomly to an image. PIV analysis was than carried out with the original image used as the first exposure and the image with noise used as the second exposure. Figure \ref{fig:error-piv} shows the values of $\sigma_{piv}$ determined from these tests using different levels of noise, and a 32 pixel x 32 pixel analysis region. 

\begin{table}
\begin{center}
\begin{tabular}{|c|c|c|c|c|}\hline
Data & $U$ & $u_{rms}$ & $\lambda$ & $Re_\lambda$ \\
set & mm/s & mm/s & mm &    \\ \hline
A1 & 100 & 3.16 & 6.69 & 24  \\
A2 & 100 & 3.02 & 6.87 & 23  \\
A3 & 100 & 2.18 & 6.46 & 16  \\
A4 & 100 & 2.26 & 8.02 & 20  \\ \hline
B1 & 125 & 4.10 & 5.37 & 25  \\
B2 & 150 & 5.10 & 5.52 & 32  \\
B3 & 175 & 6.14 & 5.42 & 38  \\
B4 & 200 & 7.26 & 5.39 & 44  \\ \hline
C1 & 100 & 1.78 & 4.83 & 10  \\
C2 & 100 & 1.55 & 4.87 & 9  \\
C3 & 100 & 0.98 & 4.73 & 5  \\ \hline
D1 & 150 & 3.10 & 5.19 & 18  \\
D2 & 150 & 1.94 & 5.38 & 12  \\
D3 & 150 & 1.90 & 5.72 & 12  \\ \hline
E1 & 175 & 3.73 & 5.35 & 22  \\
E2 & 175 & 3.18 & 5.25 & 19  \\
E3 & 175 & 2.70 & 5.98 & 18  \\
E4 & 175 & 2.54 & 5.31 & 15  \\ \hline
F1 & 200 & 4.62 & 5.18 & 27  \\
F2 & 200 & 4.00 & 5.16 & 23  \\
F3 & 200 & 4.04 & 4.76 & 22  \\ \hline
\end{tabular} 
\caption{Flow conditions for each data set}
\label{tab:flowcond}
\end{center}
\end{table}

\begin{figure}[t!]
\centering
\includegraphics[width = 120mm, angle = 0]{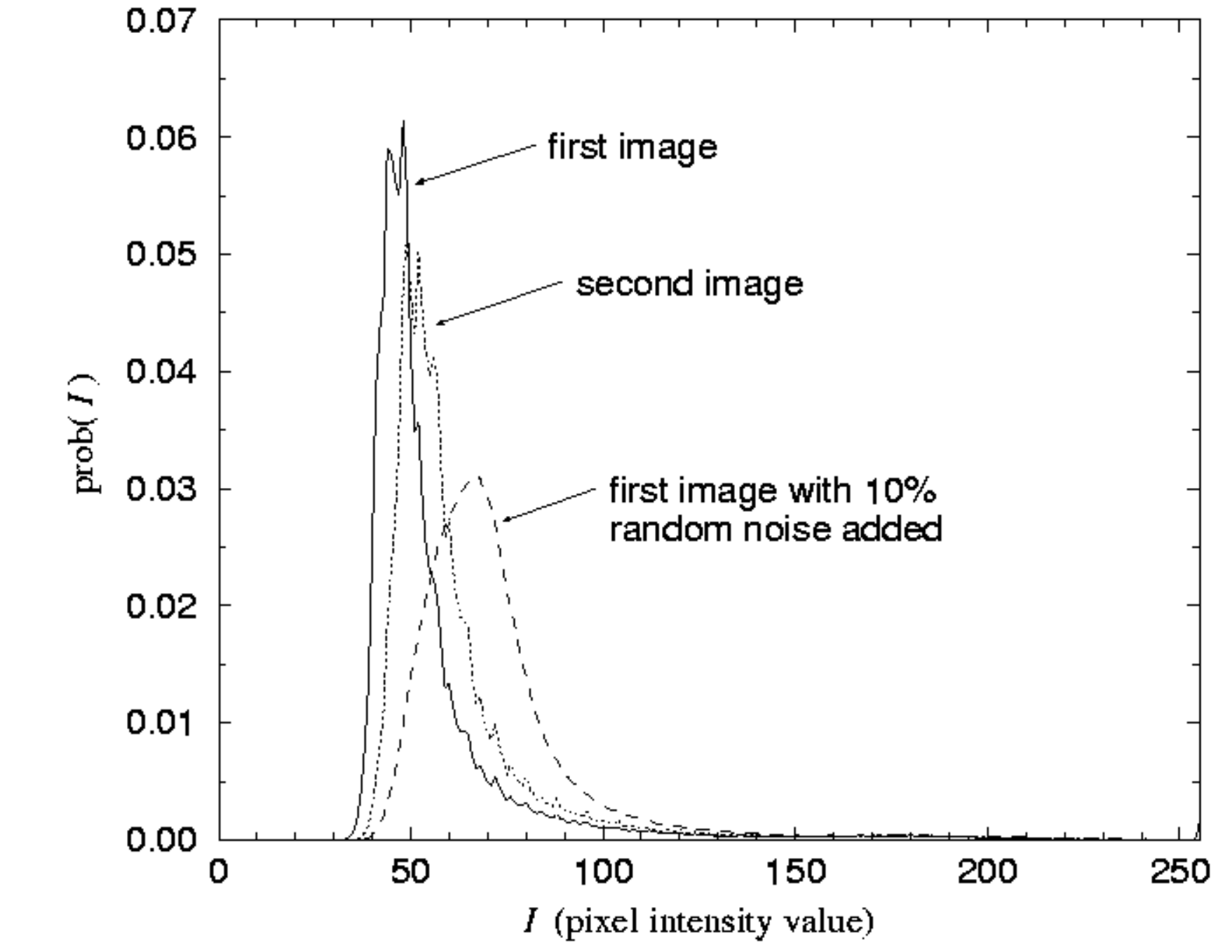}\\
\includegraphics[width = 120mm, angle = 0]{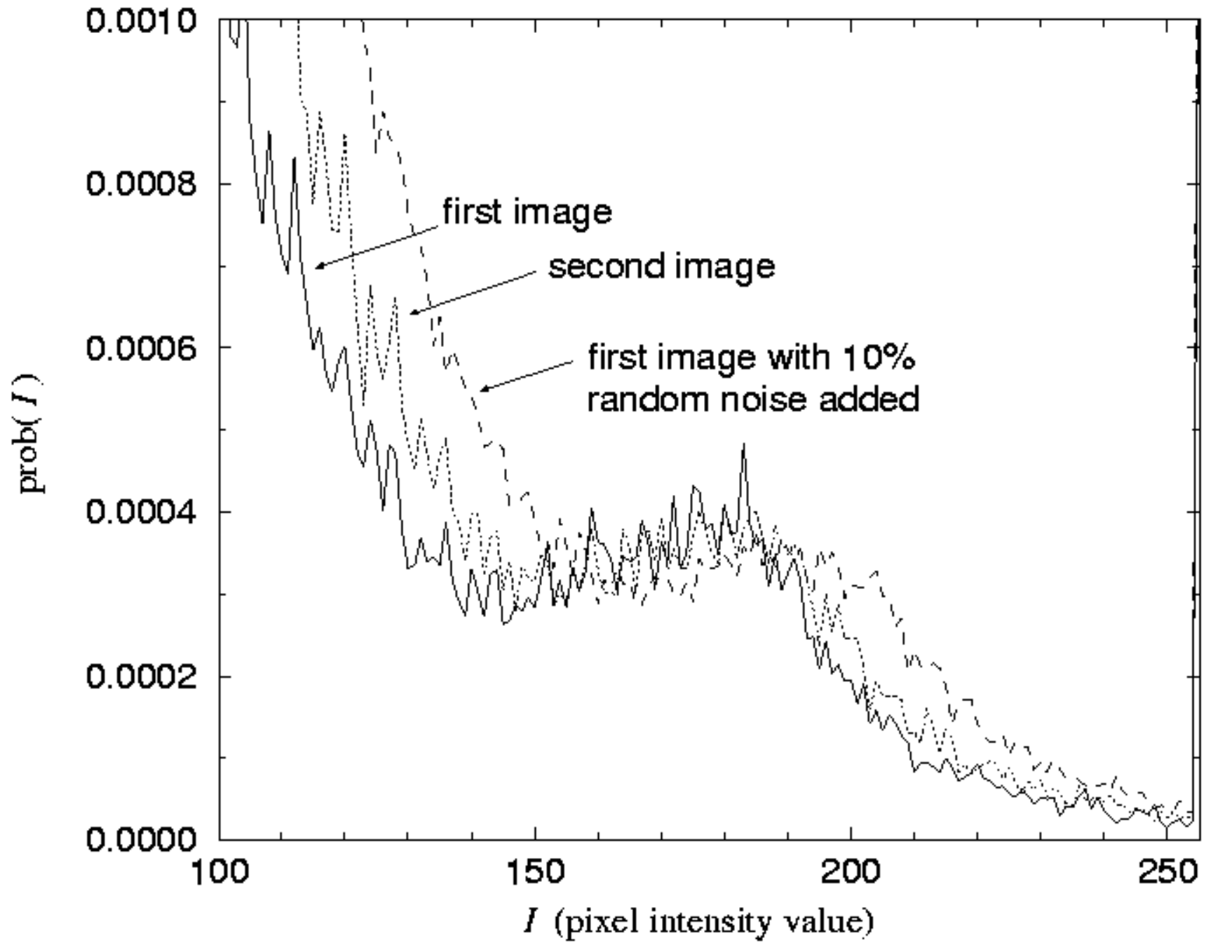} 
\caption{Probability distribution of pixel intensity values for two typical single-exposed image pairs. The first image is also shown with 10\% random noise added to each pixel. Full intensity range (top); Intensity values greater than 100 (bottom)}
\label{fig:pix-pdf-err}
\end{figure}

Figure \ref{fig:pix-pdf-err} shows the pixel intensity distribution for two typical single-exposed images of grid turbulence with a mean displacement of 160 pixels between the first and second image. The difference between the intensity distribution between the first and second images is less than the difference between the first image and the same image with 0 to 10\% random noise added. This suggests that the addition of 0 to 10\% random noise to an image will approximate the actual PIV error. Referring to Figure \ref{fig:error-piv}, a value of 0.2 pixels is an appropriate estimate of $\sigma_{piv}$. From a similar analysis for double-exposed images, a value of 0.3 pixels for $\sigma_{piv}$ was obtained (Nicolaides, 1997).

\subsection{Calculation of Velocity Gradients and Vorticity} 
The velocity gradients in the image plane ($\partial u/\partial x$, $\partial u/\partial y$, $\partial v/\partial x$ and $\partial v/\partial y$) and the vorticity component perpendicular to the image plane ($\omega_z$) are determined locally from a two dimensional polynomial that is fitted to the local velocity data. The velocity components at the grid point and the 12 nearest grid points are fitted to the following second order polynomials:

\begin{equation}
u(x,y)=u_0+u_1x+u_2y+u_3xy+u_4x^2y+u_5xy^2+u_6x^2+u_7y^2+u_8x^2y^2
\label{u-velocity}
\end{equation}
\begin{equation}
v(x,y)=v_0+v_1x+v_2y+v_3xy+v_4x^2y+v_5xy^2+v_6x^2+v_7y^2+v_8x^2y^2
\label{v-velocity}
\end{equation}
The coefficients are determined using a Chi-squared fitting procedure. The local velocity gradients are determined by analytical differentiation of equations \ref{u-velocity} and \ref{v-velocity}, and vorticity is determined according to:
\begin{equation}
\omega_z=\frac{\partial u}{\partial y}-\frac{\partial v}{\partial x}
\end{equation}

\subsection{Vorticity Error}
The total error in determining vorticity from a spatial velocity field is a combination of two factors. One source of error is the random error in the velocity field. The other source of error is a bias error caused by filtering over a finite domain in the flow in order to calculate the local velocity gradients required for the determination of vorticity. Soria and Fouras (1995) and Fouras and Soria (1998) have studied the nature of the errors and suggest a means of estimating vorticity error. Their study is based on a simulated Oseen vortex with velocity gradients calculated using a local Chi-square polynomial fit, as was done in this study. Soria and Fouras (1995) and Fouras and Soria (1998) found that the effect of the bias error was to underestimate the vorticity at regions close to the centre or core of a structure. They found that the effect of the random error was to randomly distribute measurements around the biased vorticity measurement. The measured vorticity can be expressed as:

\begin{equation}
\omega_{measured}=\omega+\omega_{bias}+\omega_{random}
\end{equation}

Soria and Fouras (1995) and Fouras and Soria (1998) provided a formalism that allowed the calculation of $\omega_{bias}$ and $\omega_{random}$ in terms of $\Delta/L$ where $\Delta$ is the spacing between velocity measurement points and $L$ is the characteristic length scale of vorticity. For an Oseen vortex, $L$ is equal to the radius at which the vorticity is 0.6 of the vorticity at the core. For vorticity calculated from velocity data obtained using a 32 pixel x 32 pixel analysis region, $\Delta/L$ was found to be approximately 0.167. The results from Soria and Fouras (1995) and Fouras and Soria (1998) show that this would give a maximum value of $\omega_{bias}$ equal to approximately -2.5\% of $\omega$ and a value of $\omega_{random}$ of approximately $\pm$ 5\% of $\omega$. For a 64 pixel x 64 pixel analysis region,  $\Delta/L$ was found to be approximately 0.333, which corresponds to a value of $\omega_{bias}$ equal to approximately -9\% of $\omega$ and a value of $\omega_{random}$ of approximately $\pm$ 6\% of $\omega$.

\subsection{Circulation}
An independent check on the accuracy of the vorticity measurement of a vortical structure was carried out by calculating another quantity that describes vortical motion \textendash \space the circulation, defined by:

\begin{equation}
\Gamma=\oint_{c}u \cdot dl
\label{circ1}
\end{equation}

where $\Gamma$ is the circulation and, in this case, $u$ is the tangential component of velocity along the contour $c$. The circulation can also be determined from:
 
\begin{equation}
\Gamma=\int_{\Omega}\omega \cdot dA
\label{circ2}
\end{equation}

where $\omega$ is the vorticity distribution within an area $\Omega$ enclosed by the contour $c$. The values determined from equation \ref{circ1} and equation \ref{circ2} should be equal.

\begin{figure}[t!]
\centering
\includegraphics[width = 120mm, angle = 0]{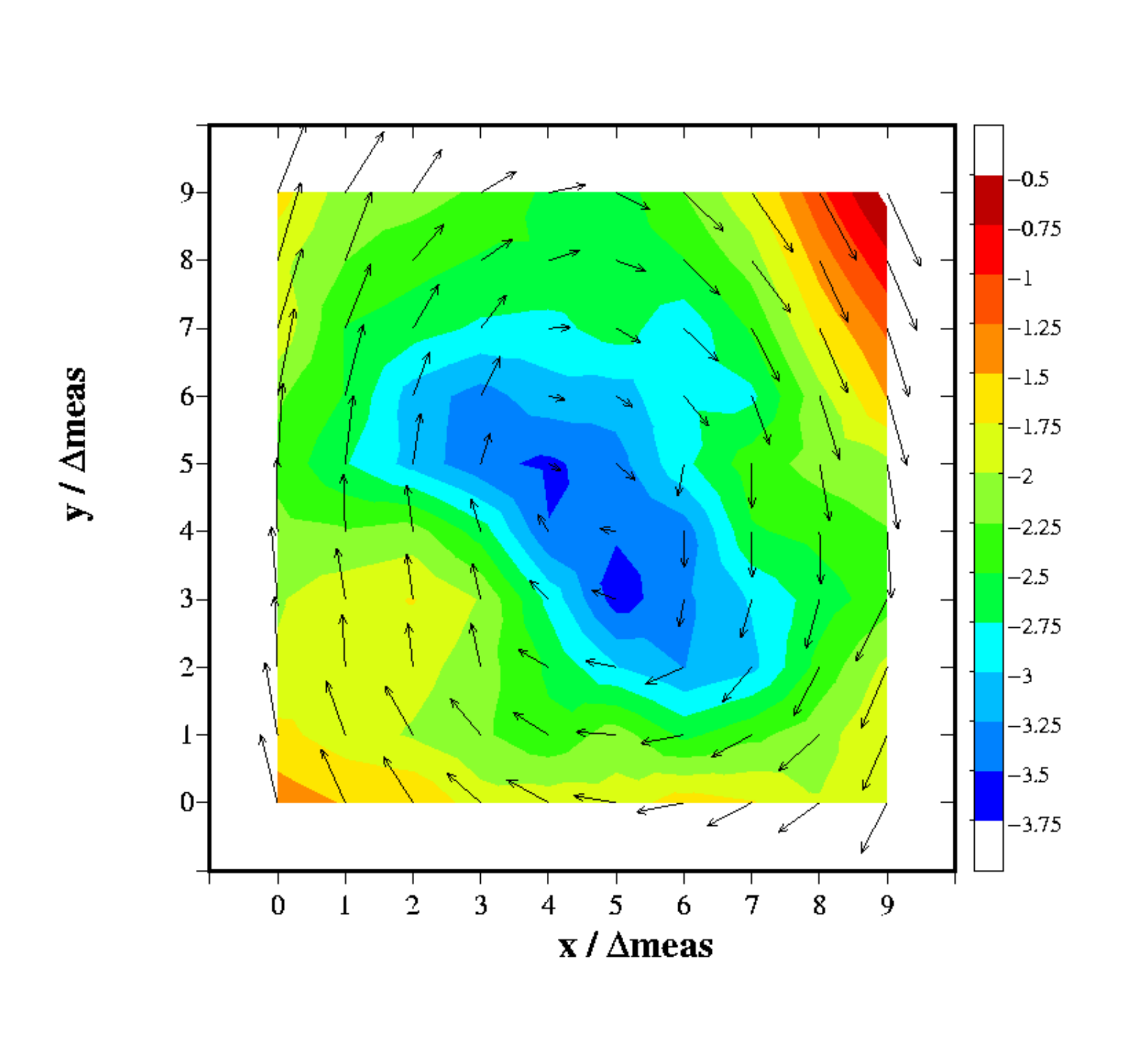}\\
\includegraphics[width = 120mm, angle = 0]{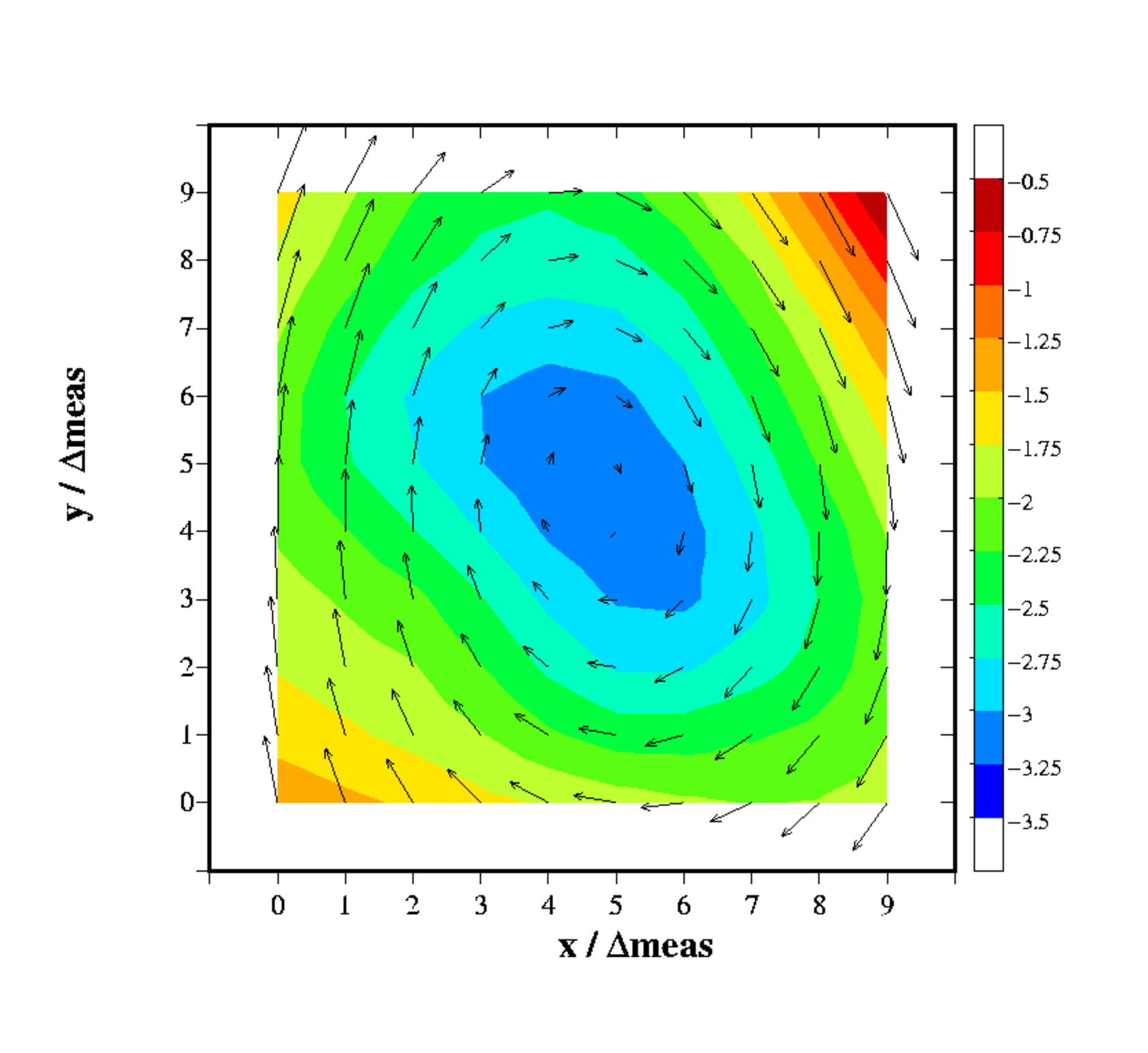} 
\caption{Local velocity and vorticity field of a structure. $Re_\lambda = 23$. Vectors indicate fluctuating velocity at each location and colour contours show $\omega/\omega '$. Raw data (top); Filtered data (bottom). Vector spacing is 16 pixels.}
\label{fig:circ}
\end{figure}

The circulation around a closed contour was calculated using both the equations \ref{circ1} and \ref{circ2} for a sub-region of a velocity field and its associated vorticity field. The sub-region contained only one apparent vortical structure. The raw field was verified and spatially filtered and circulation calculations were performed using both the raw and filtered data. The sub-region, with both raw and filtered data, is shown in Figure \ref{fig:circ}. The results of the calculation are presented in Table \ref{tab:circ}. In all cases, the value of the vorticity area integral was less than 3\% less than the equivalent velocity line integral, which is within the range of underestimation of vorticity predicted in the previous section.

\begin{table}
\begin{center}
\begin{tabular}{|c|c|c|c|}\hline
Calculation region & Circulation & Circulation & Difference  \\ 
$x/\Delta meas x y/\Delta meas$ & Velocity line integral & Vorticity area integral & \% \\ \hline
10 x 10 - raw & 144.7 & 140.9 & -2.6 \\
10 x 10 - smooth & 139.4 & 137.0 & -1.7 \\ \hline
6 x 6 - raw & 53.7 & 53.3 & -0.8 \\
6 x 6 -smooth & 52.2 & 50.8 & -2.7 \\
\hline
\end{tabular} 
\caption{Comparison of circulation calculated using the velocity line integral and vorticity area integral. All circulation values obtained are normalised by $\mathbf{u}\lambda$. Vector spacing is 16 pixels; $\mathbf{u}=3.02$ mm/s; $\lambda = 6.9$ mm}
\label{tab:circ}
\end{center}
\end{table}

\section{Results}
\subsubsection{Isotropy of flow}
Before continuing with the investigation of this flow, it is important that the isotropy of the flow be verified. This can be done by investigating the relationship between the velocity gradients, as described in Goldstein (1965), Comte-Bellot and Corrsin (1971), and Hinze (1975), and is summarised by the following four equations:
\begin{equation}
\overline{(\frac{du}{dx})^2}=\overline{(\frac{dv}{dy})^2}
\end{equation}
\begin{equation}
\overline{(\frac{du}{dy})^2}=\overline{(\frac{dv}{dx})^2}
\end{equation}
\begin{equation}
\overline{(\frac{du}{dx})^2}=\overline{\frac{1}{2}(\frac{du}{dy})^2}
\end{equation}
\begin{equation}
\overline{(\frac{dv}{dy})^2}=\overline{\frac{1}{2}(\frac{dv}{dx})^2}
\end{equation}

These relationships were found to be satisfied to within 30\% for most of the experimental conditions investigated (Nicolaides, 1997). Where these conditions were not satisfied to within 30\% it was observed that the background turbulence intensity was also at the higher end of the range (Nicolaides, 1997). Tsinober er.al. (1992) measured the same gradients, and their results indicate that the gradient relations are satisfied to within 40\%. The value of $u_{rms}$ is generally higher than the value of $v_{rms}$ in grid turbulence due to the effect of the convective velocity. In these experiments the ratio $u_{rms}/v_{rms}$ is between 1.05 and 1.20, while for Tsinober et.al. (1992) the ratio was typically 1.2. This anisotropy is probably the main contributing factor to the discrepancies found in satisfying the isotropy conditions. 

\subsection{Instantaneous Velocity and Vorticity}
Instantaneous velocity fields are determined directly from the pair of particle images analysed using PIV. The instantaneous fluctuating velocity fields are obtained by subtracting the mean velocity field calculated from an ensemble of instantaneous velocity fields from each of the individual velocity fields according to the equation:
\begin{equation}
u(x,y)=U(x,y)-\bar{U}(x,y)
\end{equation}
where the overbar indicates the mean value. The equation shown is for the $u$ component of velocity in the $x$ direction, but a similar equation can be written for the $v$ component of velocity in the $y$ directions.

\begin{figure}[t!]
\centering
\includegraphics[width = 150mm, angle = 0]{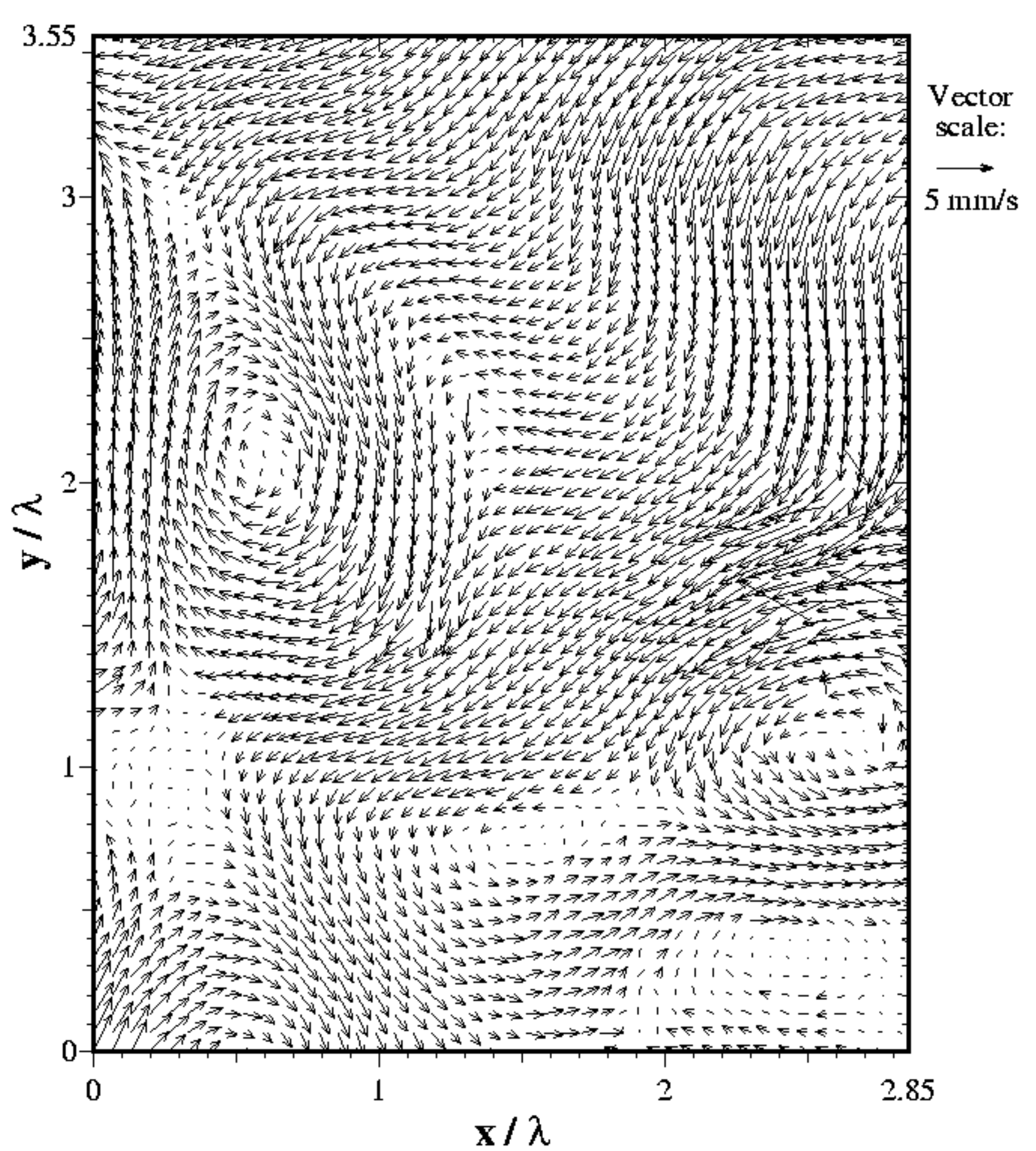}
\caption{Typical fluctuating velocity field \textendash \space raw. $Re_\lambda = 23$. Image pair from data set A2. Velocity field calculated using a 32 pixel x 32 pixel analysis window with 0.5 overlap.}
\label{fig:flucvelraw}
\end{figure}

\begin{figure}[t!]
\centering
\includegraphics[width = 150mm, angle = 0]{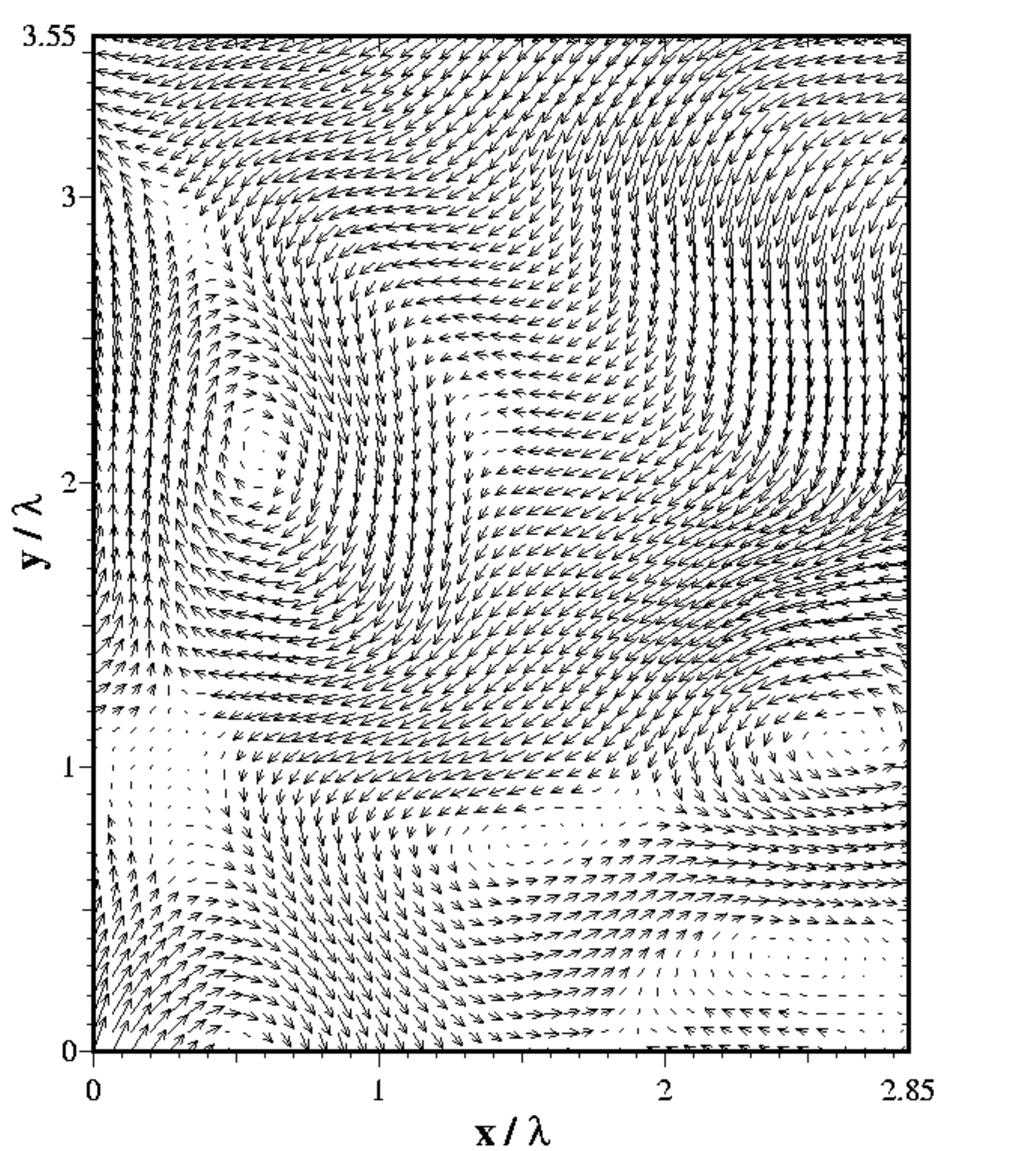}
\caption{Filtered fluctuating velocity field corresponding to the raw fluctuating velocity field shown in Figure \ref{fig:flucvelraw}. $Re_\lambda = 23$. Image pair from data set A2. Velocity field calculated using a 32 pixel x 32 pixel analysis window with 0.5 overlap.}
\label{fig:flucvelsm}
\end{figure}

A typical fluctuating velocity field obtained from cross-correlation analysis of two single-exposed images is shown in Figure \ref{fig:flucvelraw}. Vortical structures appear to be clearly resolved and of the order of the Taylor scale of the flow ($\lambda$). The raw field was verified and spatially filtered (Figure \ref{fig:flucvelsm}). In comparison to the raw field, the filtered field shows structures that are more easily identified. Irrespective of the filtering, the structures appear to be quantitatively and qualitatively the same in both the raw and filtered velocity fields.

\begin{figure}[t!]
\centering
\includegraphics[width = 80mm, angle = 0]{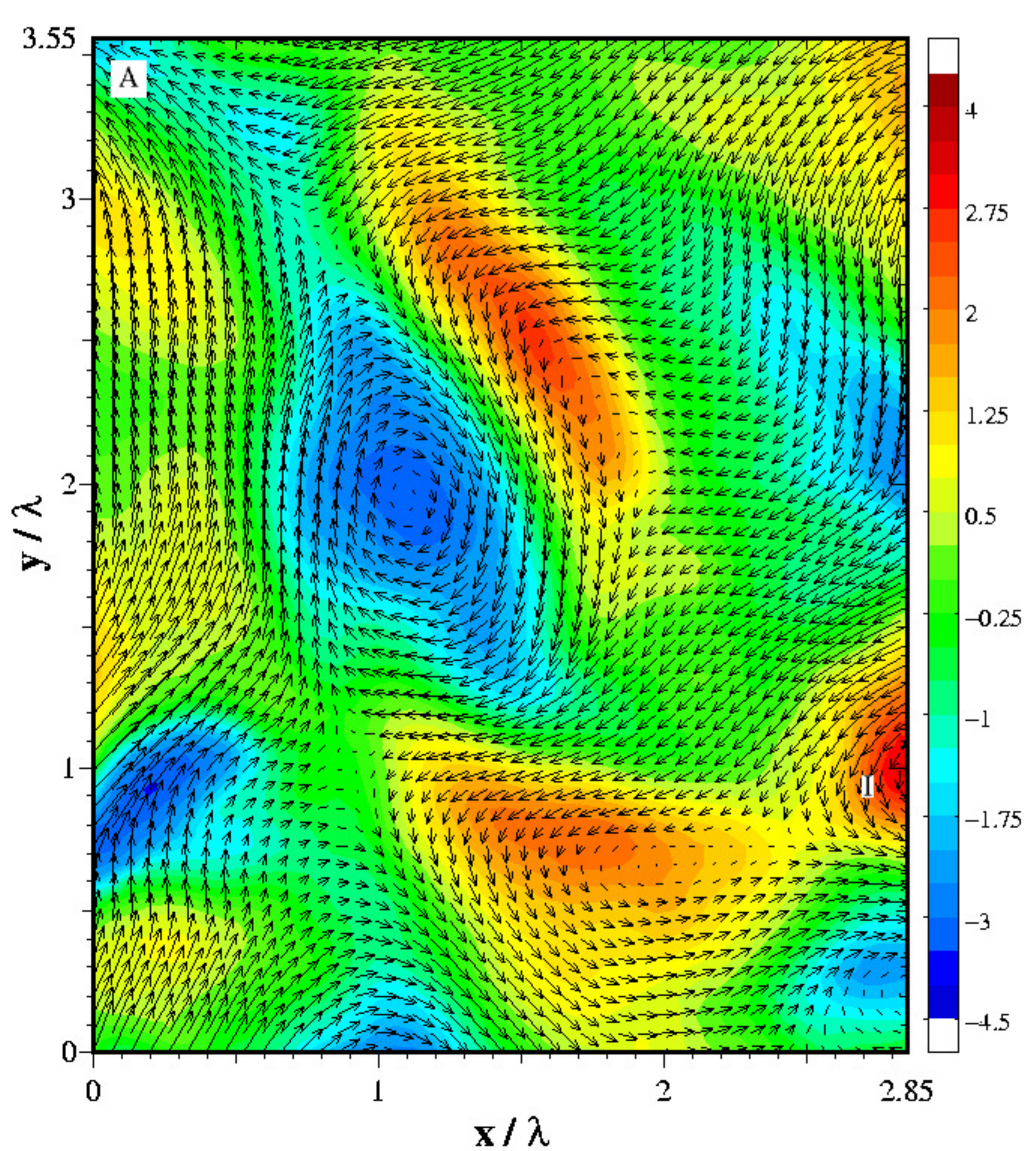}
\includegraphics[width = 80mm, angle = 0]{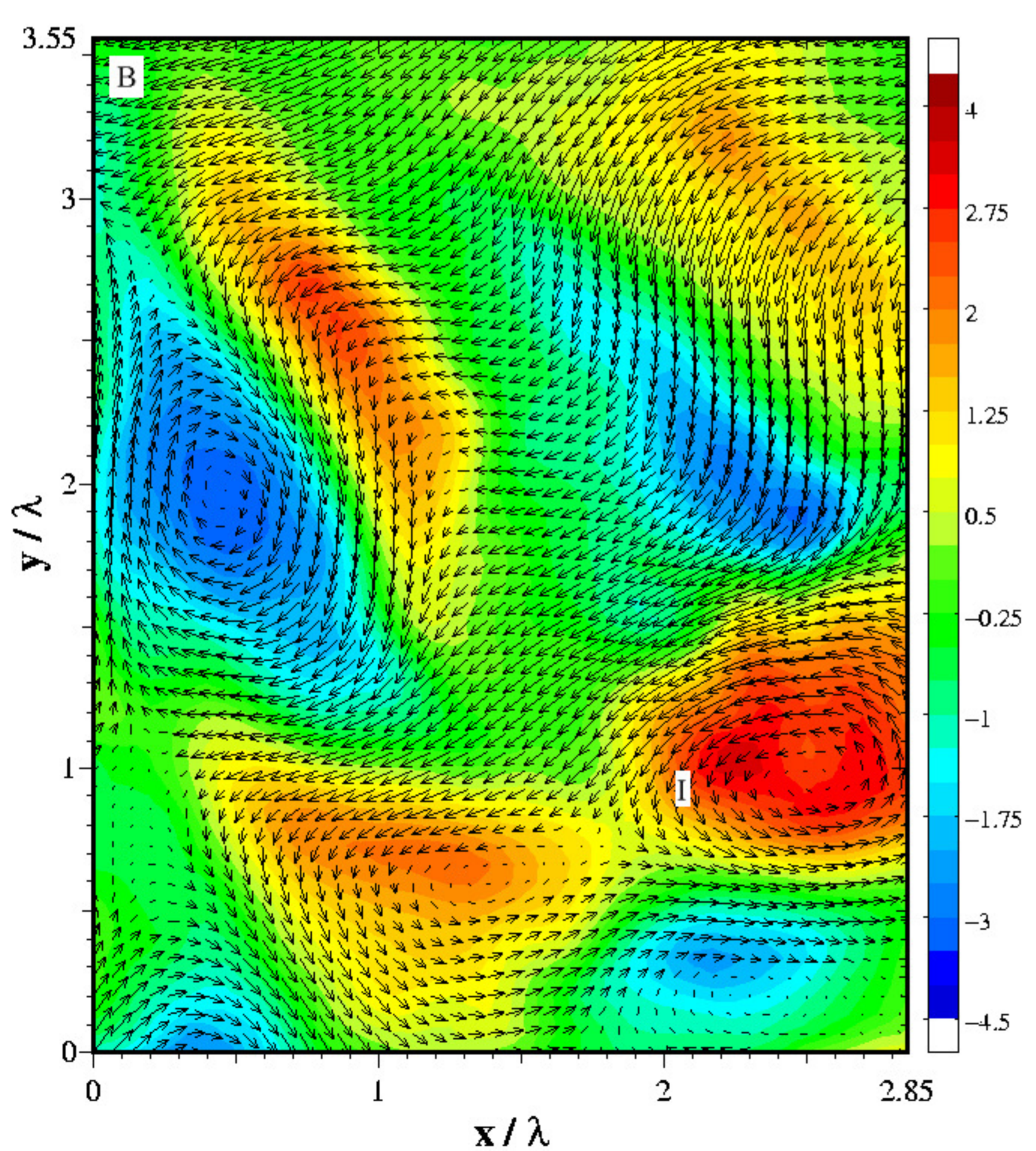} \\
\includegraphics[width = 80mm, angle = 0]{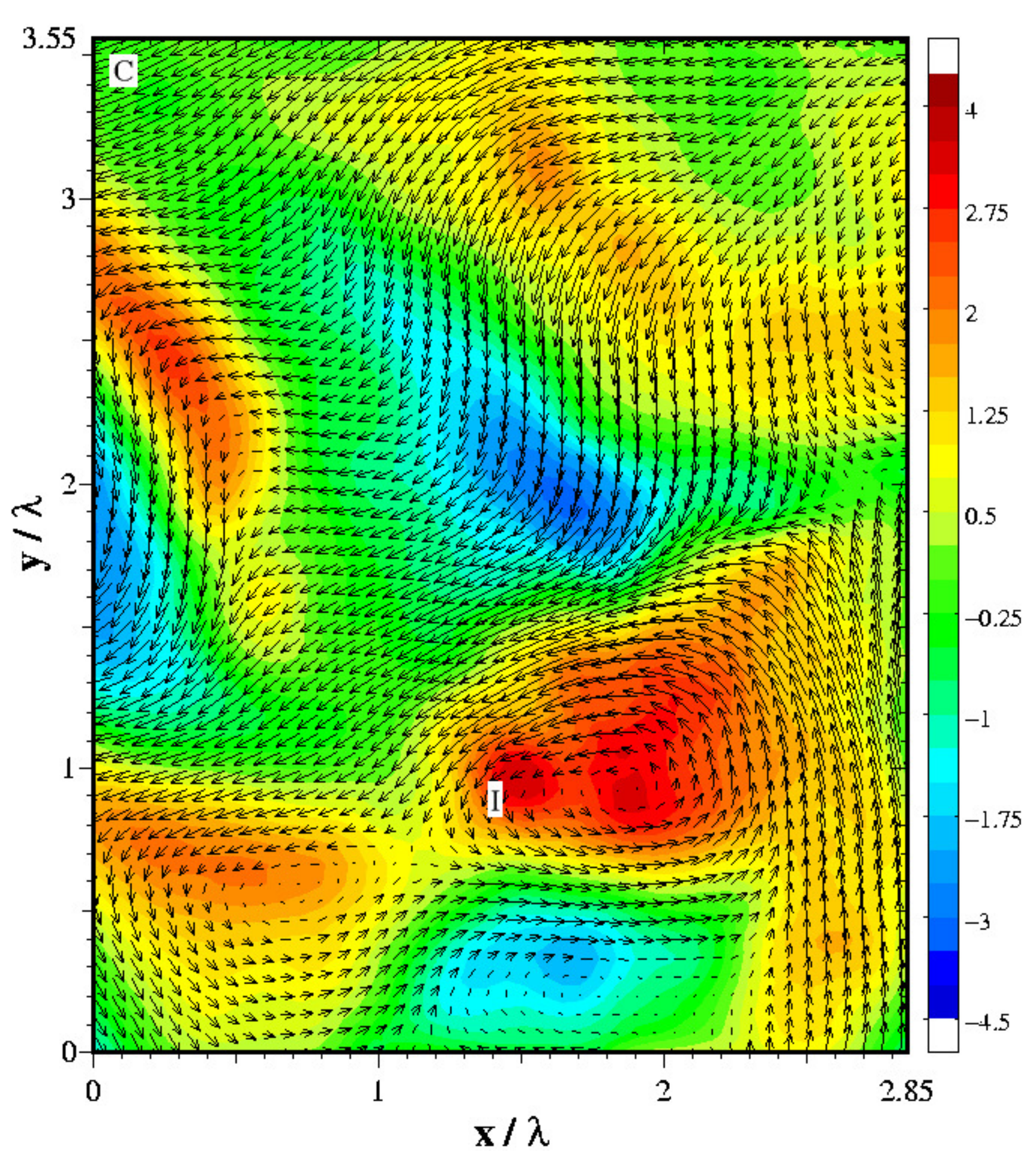}
\includegraphics[width = 80mm, angle = 0]{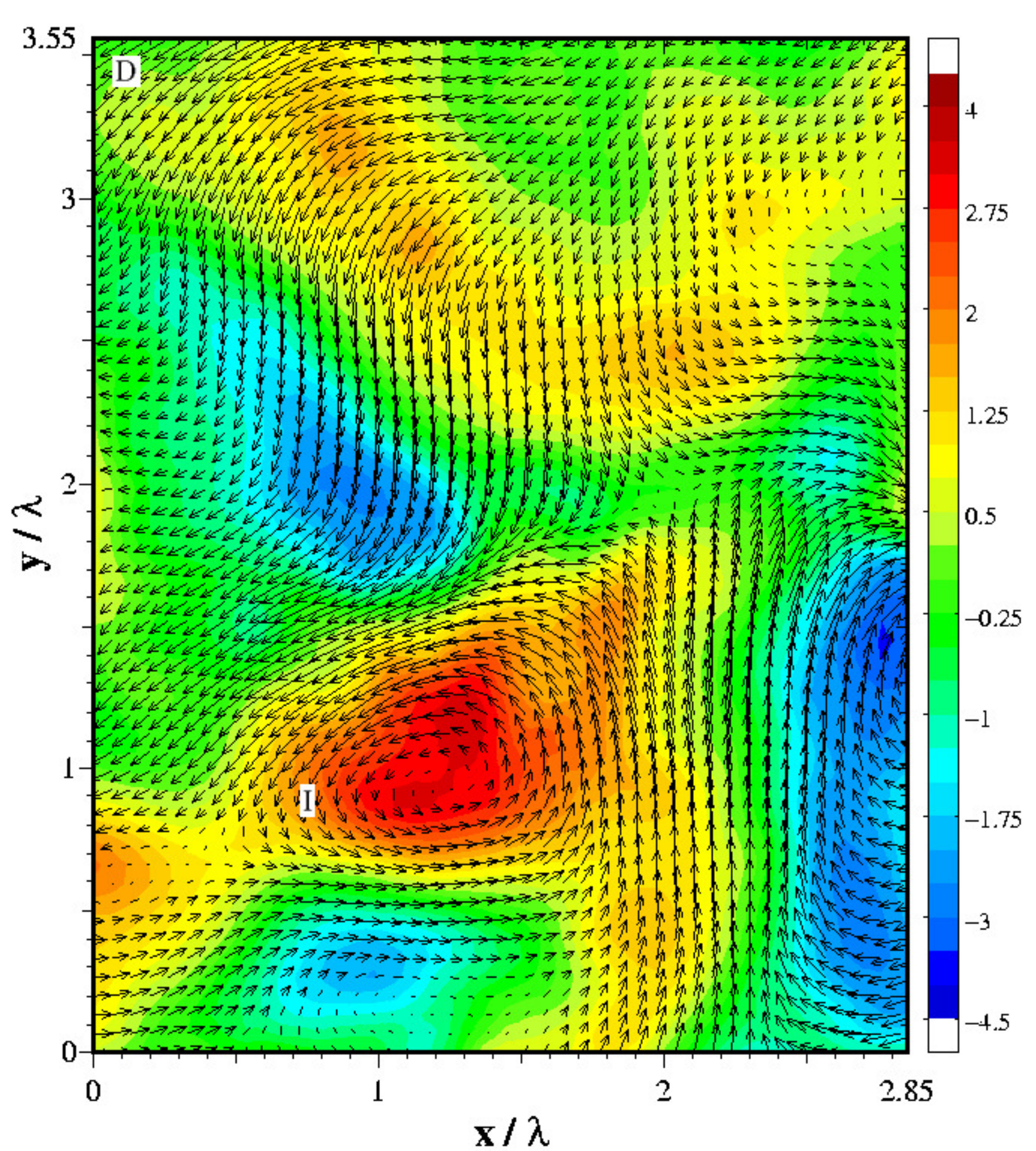}
\caption{Time series of filtered fluctuating velocity and vorticity. $Re_\lambda = 23$. Image pair from data set A2. Contour plots of vorticity show $\omega/\omega'$. Flow displacement between frames is approximately $0.65\lambda$. ($\lambda\approx$ 6.9 mm)}
\label{fig:vortdev}
\end{figure}

Four fields of filtered fluctuating velocity and vorticity are shown in Figure \ref{fig:vortdev}. These fields constitute a time series with the mean flow displacement between each field being approximately 0.65$\lambda$. In each consecutive image from A to D the progression of the structures across the image plane can be seen. The progression of one structure identified by 'I' is indicated. 

\begin{figure}[t!]
\centering
\includegraphics[width = 80mm, angle = 0]{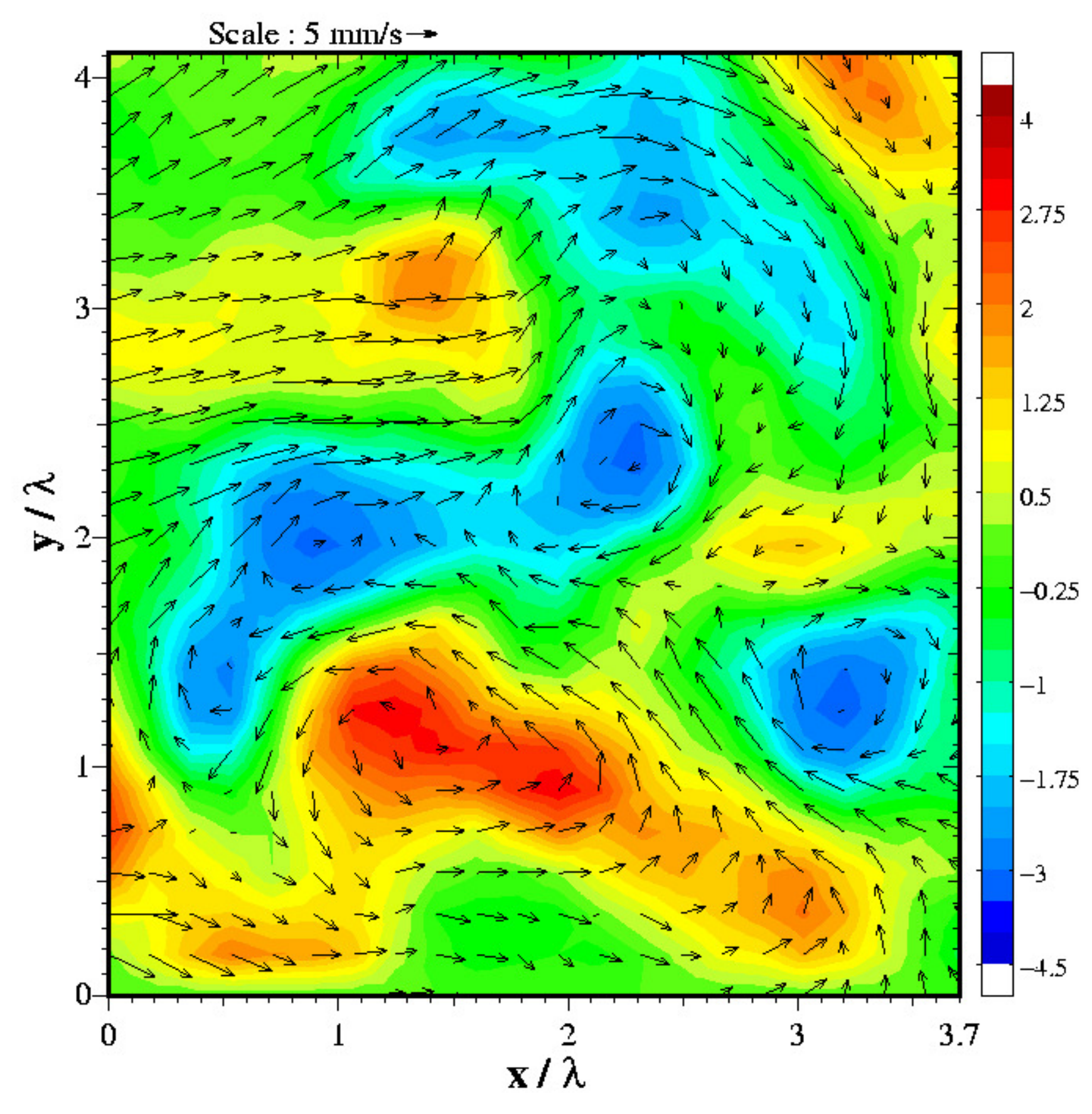}
\includegraphics[width = 80mm, angle = 0]{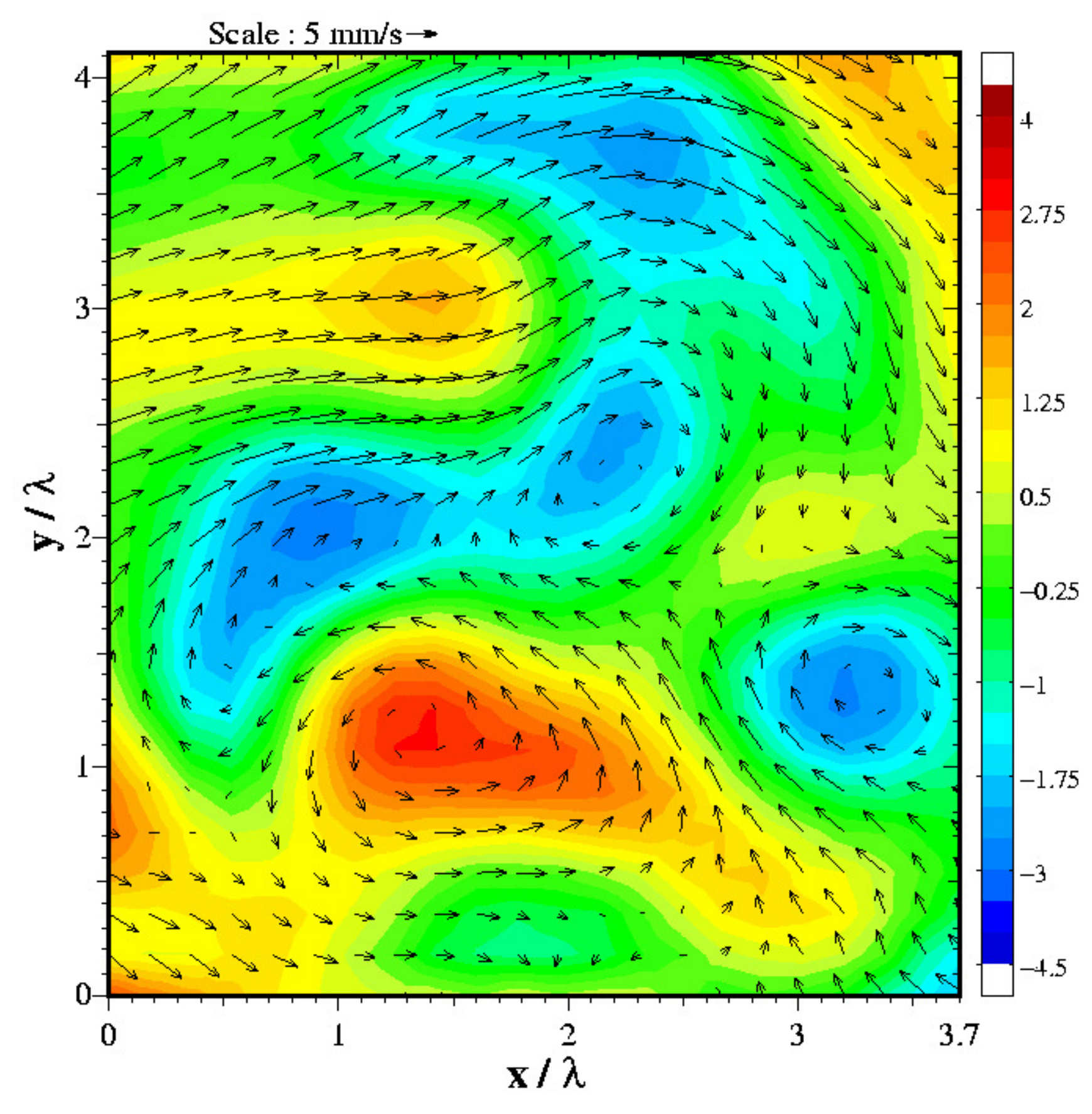} 
\caption{Typical fluctuating velocity field. $Re_\lambda = 44$. Double-exposure image from data set B4. Contour plots of vorticity show $\omega/\omega'$. Raw data (left); Filtered data (right). ($\lambda\approx$ 5.4 mm)}
\label{fig:flucvel-b4}
\end{figure}

Figure \ref{fig:flucvel-b4} shows the fluctuating velocity and the corresponding vorticity at a higher Taylor-Reynolds number ($Re_\lambda = 44$) determined from a double-exposed image. Structures with magnitude of order $\lambda$ appear to be resolved, but with fewer vectors than in the previous result (Figure \ref{fig:flucvelraw} and Figure \ref{fig:flucvelsm}). The spatial filtering appears to have a greater impact on the velocity field at this lower resolution, however the size and shape of the structures in the field are still faithfully maintained.

\subsubsection{Decay of Turbulence}
\begin{figure}[t!]
\centering
\includegraphics[width = 120mm, angle = 0]{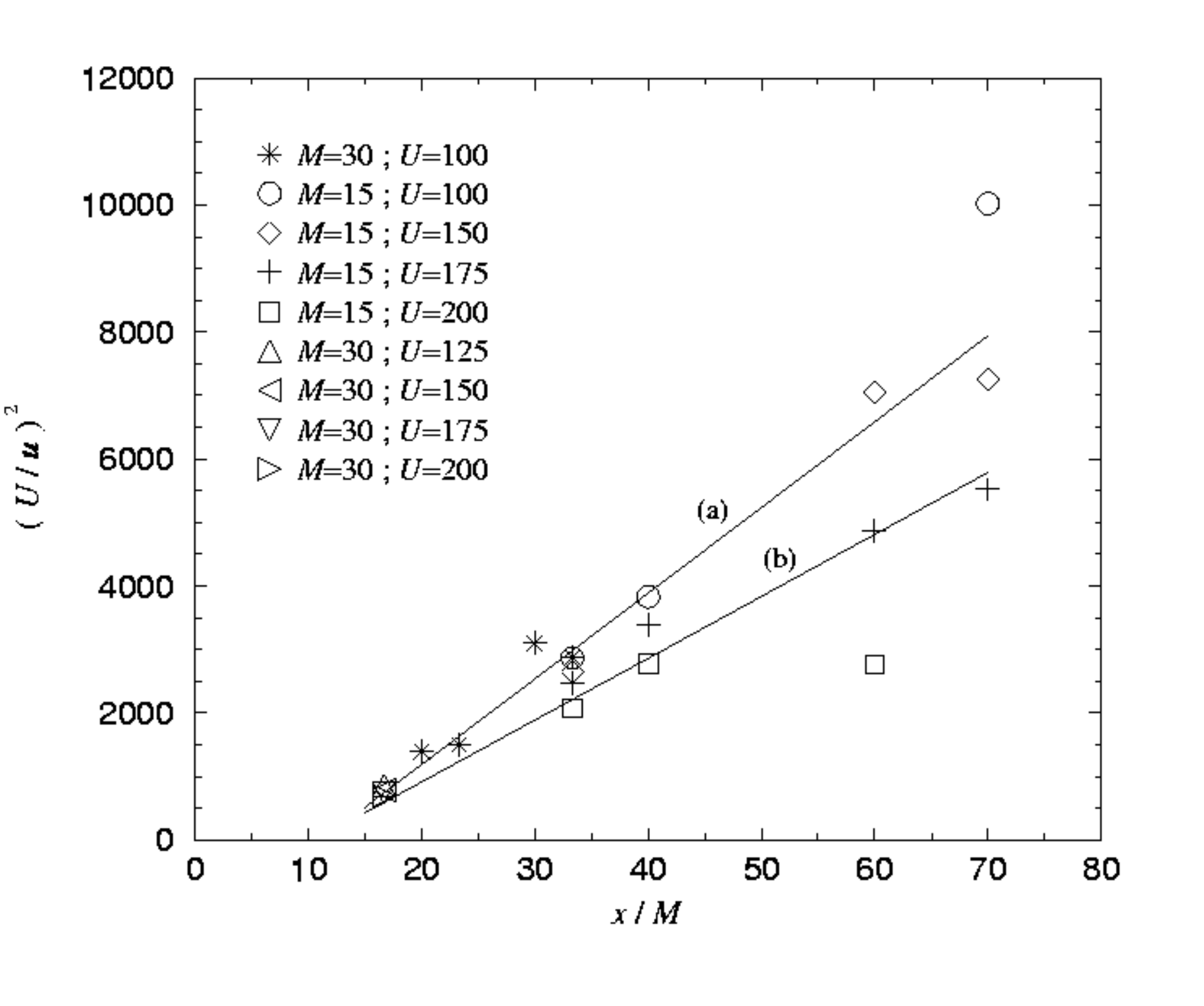}
\caption{Decay of turbulence for experimental conditions. (a) line fitted to data for $U=100$, 125 and 150 mm/s; (b) line fitted to data for $U=175$ and 200 mm/s, excluding data point for $M=30$, $U=200$ mm/s, $x/M=60$}
\label{fig:virtorig}
\end{figure}

In order to determine the virtual origin of decay in the experimental apparatus, the decay of turbulence dowstream from the grid was plotted.  The value of $(U/\mathbf{u})^2$ was plotted against the distance $x/M$, as shown in Figure \ref{fig:virtorig}. $U$ is the mean velocity in the $x$ direction and $\mathbf{u}$ is the root mean square of the turbulent velocity fluctuations, defined by:

\begin{equation}
{\bf{u}}=\sqrt{\frac{1}{2}(\langle{u_{rms}^2}\rangle + \langle{v_{rms}^2}\rangle)}
\end{equation}

Previous measurements of grid turbulence have suggested that the decay of turbulent energy is approximately linear in the initial period up to $x/M=100$ (Batchelor and Townsend (1947), Mohamed and LaRue (1990), Snyder and Lumley (1971)), and this appears to be the case for the results presented in Figure \ref{fig:virtorig}. However, there is insufficient data to determine the decay function for each experimental condition, so the line (a) shown in Figure \ref{fig:virtorig} is fitted to the data for $U=100$, 125 and 150 mm/s and the line (b) is fitted to the data for $U=175$ and 200 mm/s, excluding the data for the case where $M=15$, $U=200$ and $x/M=60$, as it was observed to deviate from the bulk of the data. This deviation can be explained by the high levels of background turbulence recorded for this experimental condition.

The equation of line (a) is:

\begin{equation}
(\frac{U}{\bf{u}})^2=135(\frac{x}{M}-11)
\label{eqna}
\end{equation}

and for line (b) is:

\begin{equation}
(\frac{U}{\bf{u}})^2=97(\frac{x}{M}-10)
\label{eqnb}
\end{equation}

Depending on the flow conditions the virtual origin appears to be between $x/M=10$ and $x/M=11$. This is consistent with previous results (Batchelor and Townsend (1947), Snyder and Lumley (1971)), which showed that the virtual origin is dependent on the experimental conditions, but consistently lies between $x/M=10$ and $x/M=20$.

Fitting data from Batchelor and Townsend (1948), determined using two grids with $M=25.4$ mm and $M=50.8$ mm, the following equation in the same format as equations \ref{eqna} and \ref{eqnb} is obtained:

\begin{equation}
(\frac{U}{\bf{u}})^2=140(\frac{x}{M}-10)
\label{eqnbandt}
\end{equation}

This equation compares favourably to the equations \ref{eqna} and \ref{eqnb} obtained in this experimental study.

\subsubsection{Probability Distribution Functions}
Probability Distribution Functions (PDFs) are calculated for both components of the fluctuating velocity. The PDFs are calculated using all the velocity vectors in the dataset. All PDFs are calculated within $\pm 6 \sigma$ (where $\sigma$ is the standard deviation) using 90 bins, giving a bin size of 0.13$\sigma$. Each PDF is normalised so that $\sigma=1$ and the area under the curve is equal to one. The vertical axis is plotted in log scale.

\begin{figure}[t!]
\centering
\includegraphics[width = 120mm, angle = 0]{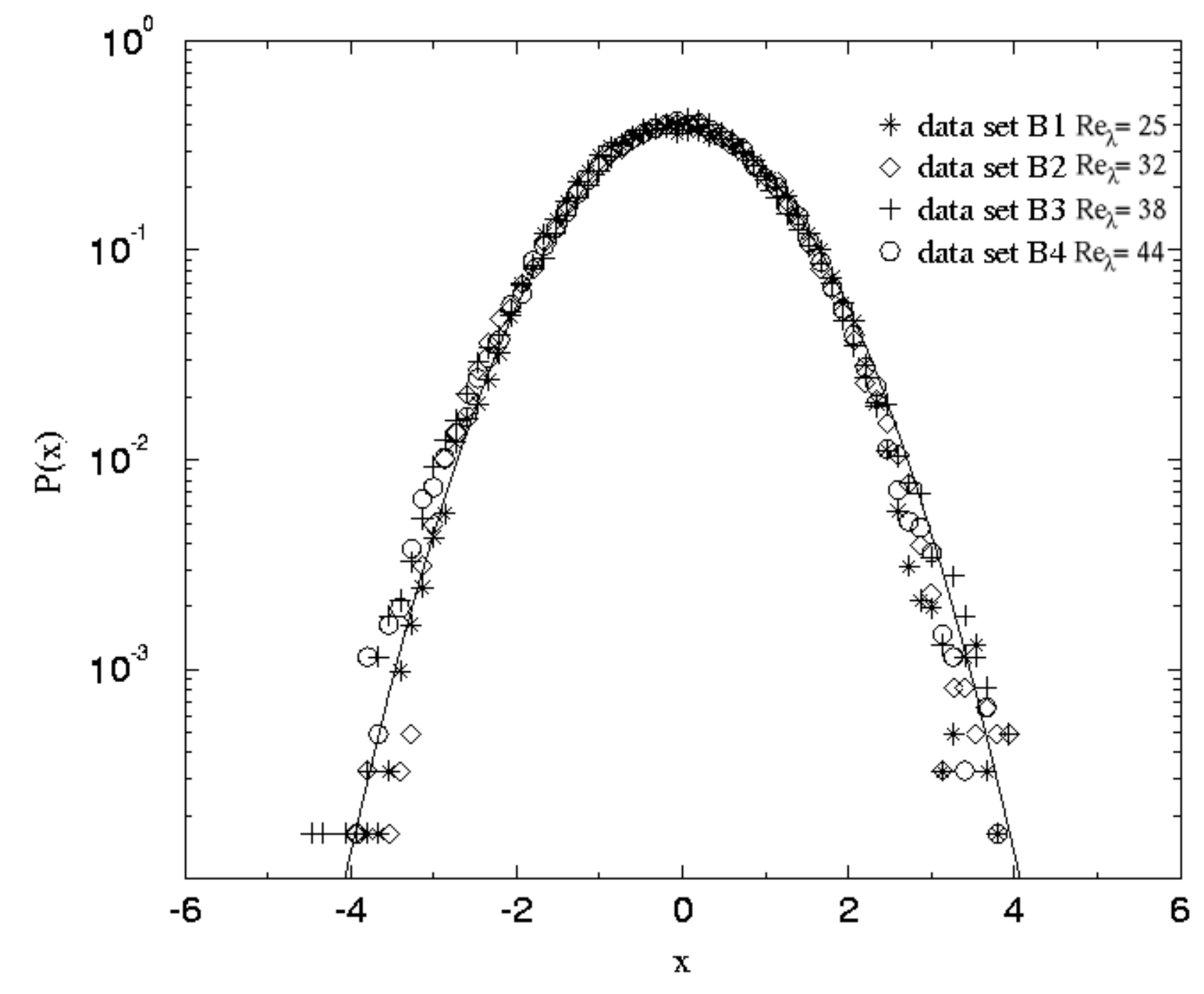}\\
\includegraphics[width = 120mm, angle = 0]{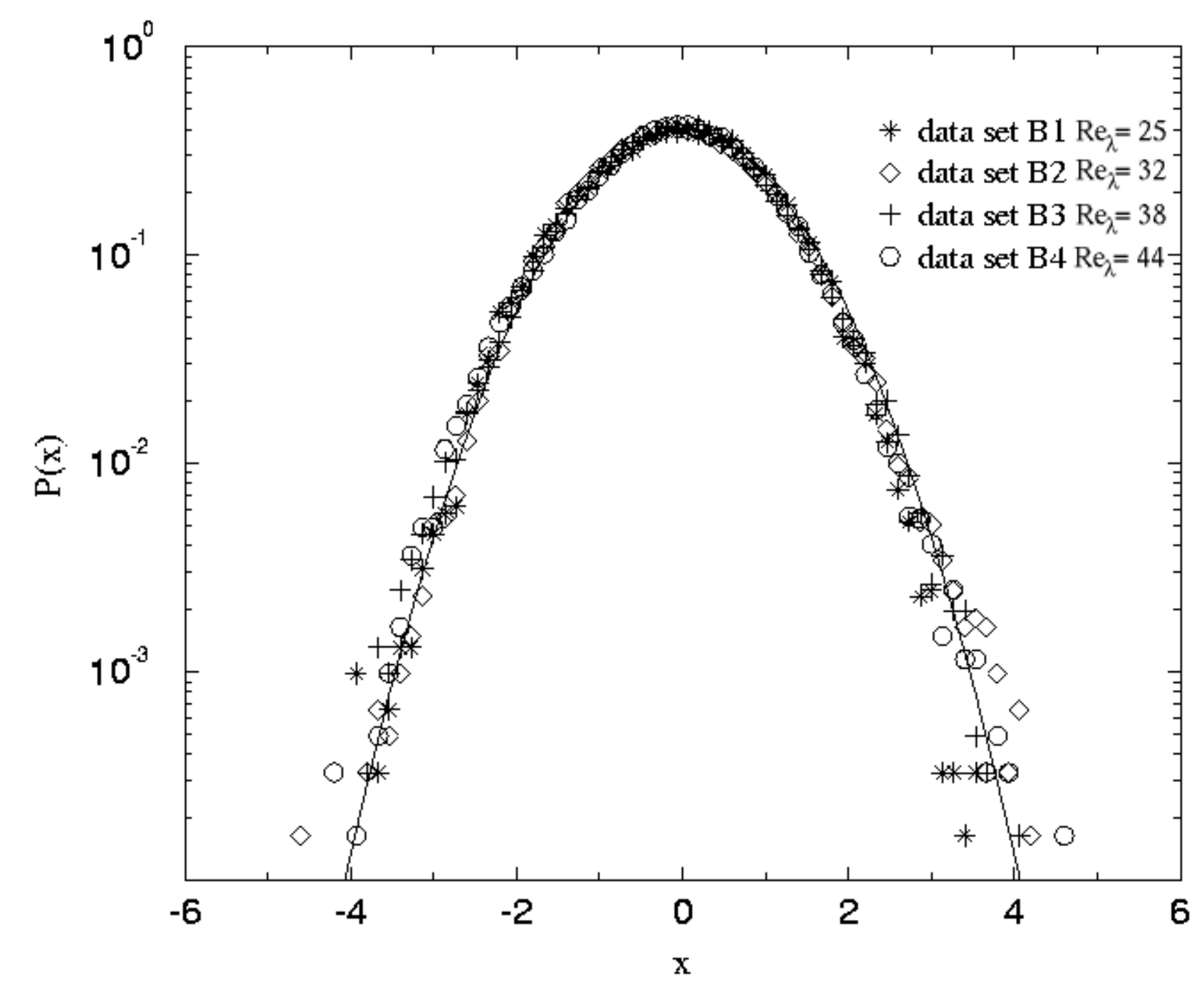} 
\caption{Probability distribution functions of fluctuating velocity components for data set B, normalised so that $\sigma=1$. The range of $Re_\lambda$ represented is from 25 to 44. The Gaussian distribution is shown as a solid line. $u$ (top); $v$ (bottom)}
\label{fig:fv-pdf-B}
\end{figure}

The PDFs of the fluctuating components of velocity for data set B, $Re_\lambda$ between 25 and 44, are shown in Figure \ref{fig:fv-pdf-B}. The distribution is close to Gaussian for all the data sets. Statistical moments were calculated for these distributions. The skewness values were close to zero and the kurtosis values were close to 3, both of which are consistent with a Gaussian distribution. Similar results were found for experimental data by Townsend (1947), and for direct numerical simulation (DNS) data by Vincent and Meneguzzi (1991). Mohamed and LaRue (1990) found that skewness increased with Reynolds number at a fixed downstream location. No apparent trends in either skewness or kurtosis were observed for the fluctuating velocity in this study. 

\subsubsection{Autocorrelation Functions}
Spatial autocorrelation functions of velocity are calculated in both the longitudinal and transverse directions determined with reference to the velocity direction. The spatial autocorrelation function, $R$, for the $u$ component of velocity in direction $x$ (the longitudinal autocorrelation of the velocity in the $x$ direction) is defined as:
\begin{equation}
R_{xx}(r)=\frac{\langle u(x,t)\cdot u(x+r,t)\rangle}{\langle u^2 \rangle}
\end{equation}
where $r$ is the spatial distance between measurement points.

\begin{figure}[t!]
\centering
\includegraphics[width = 120mm, angle = 0]{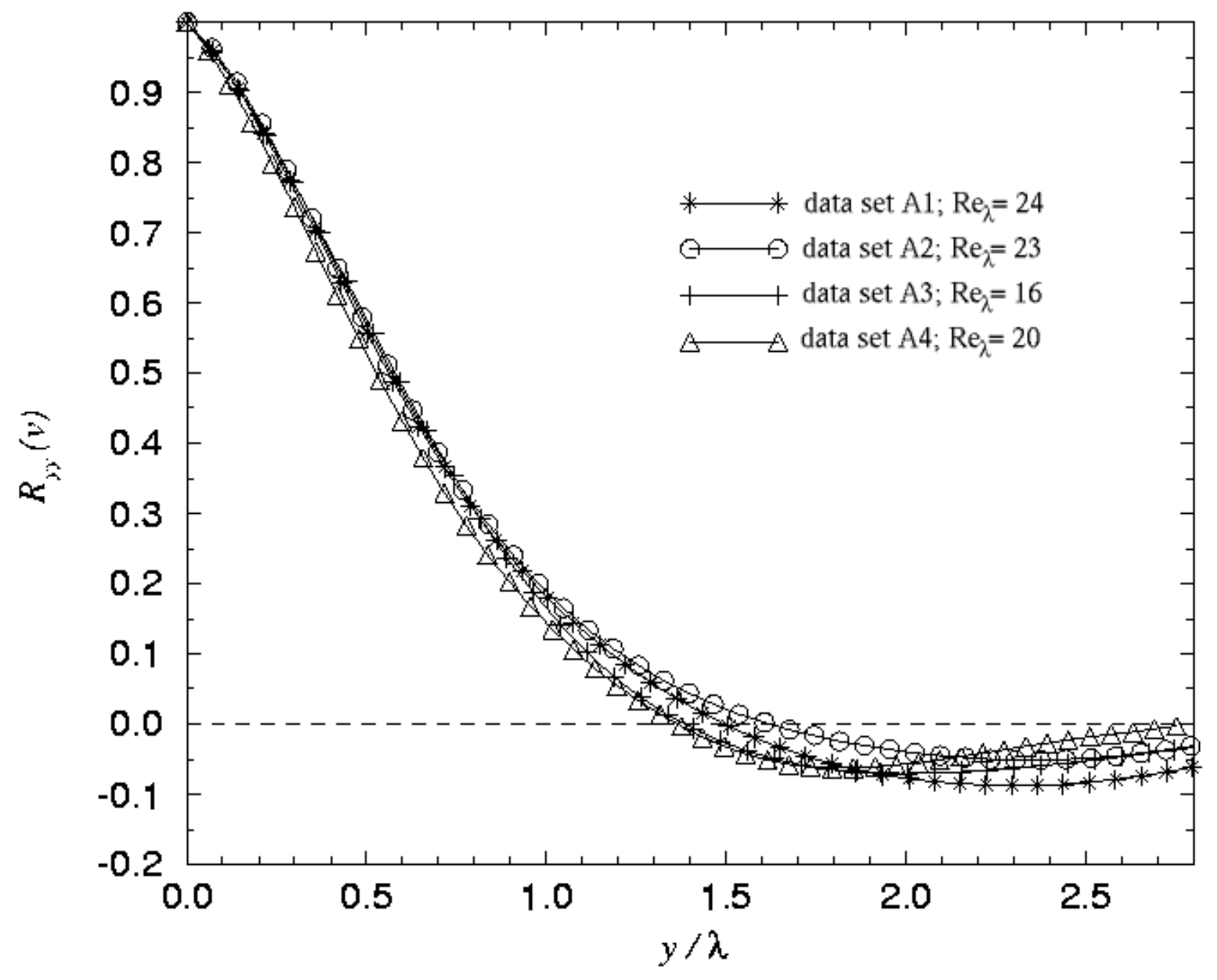} 
\caption{Longitudinal autocorrelation of $v$ for data set A. $Re_\lambda$ between 16 and 24.}
\label{fig:auto-long-vel-A}
\end{figure}

The longitudinal autocorrelation functions for $v$ for data set A, i.e $R_{yy}(v)$ for $Re_\lambda$ between 16 and 24, are shown in Figure \ref{fig:auto-long-vel-A}. Each function represents a different downstream location ($x_2$, $x_3$ etc.). The four functions appear to be quite similar and therefore independent of $Re_\lambda$, particularly at low values of $r$. The value of $r$ is normalised by $\lambda$, estimated using equation \ref{taylor-1}.

\begin{figure}[t!]
\centering
\includegraphics[width = 120mm, angle = 0]{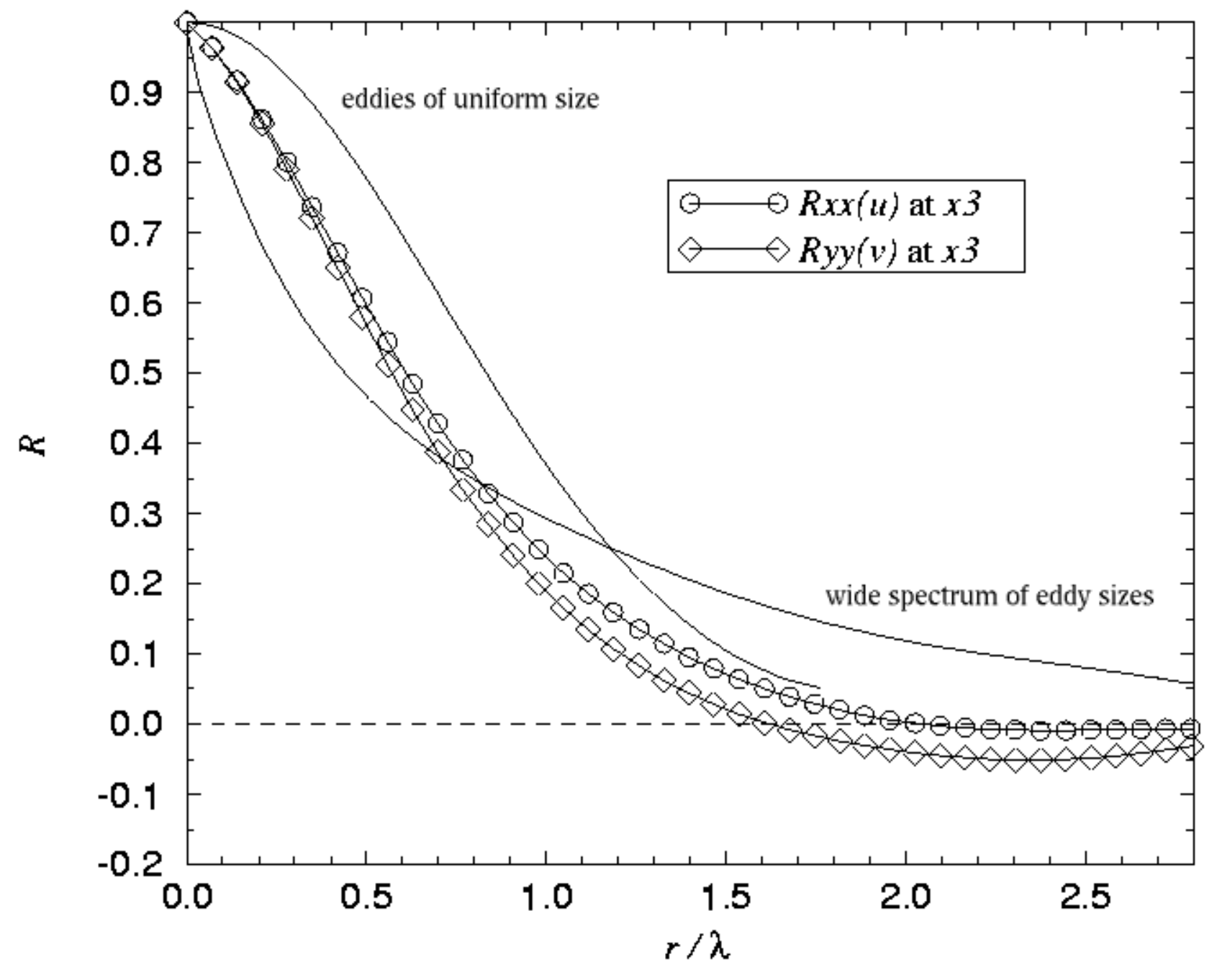} \\
\includegraphics[width = 120mm, angle = 0]{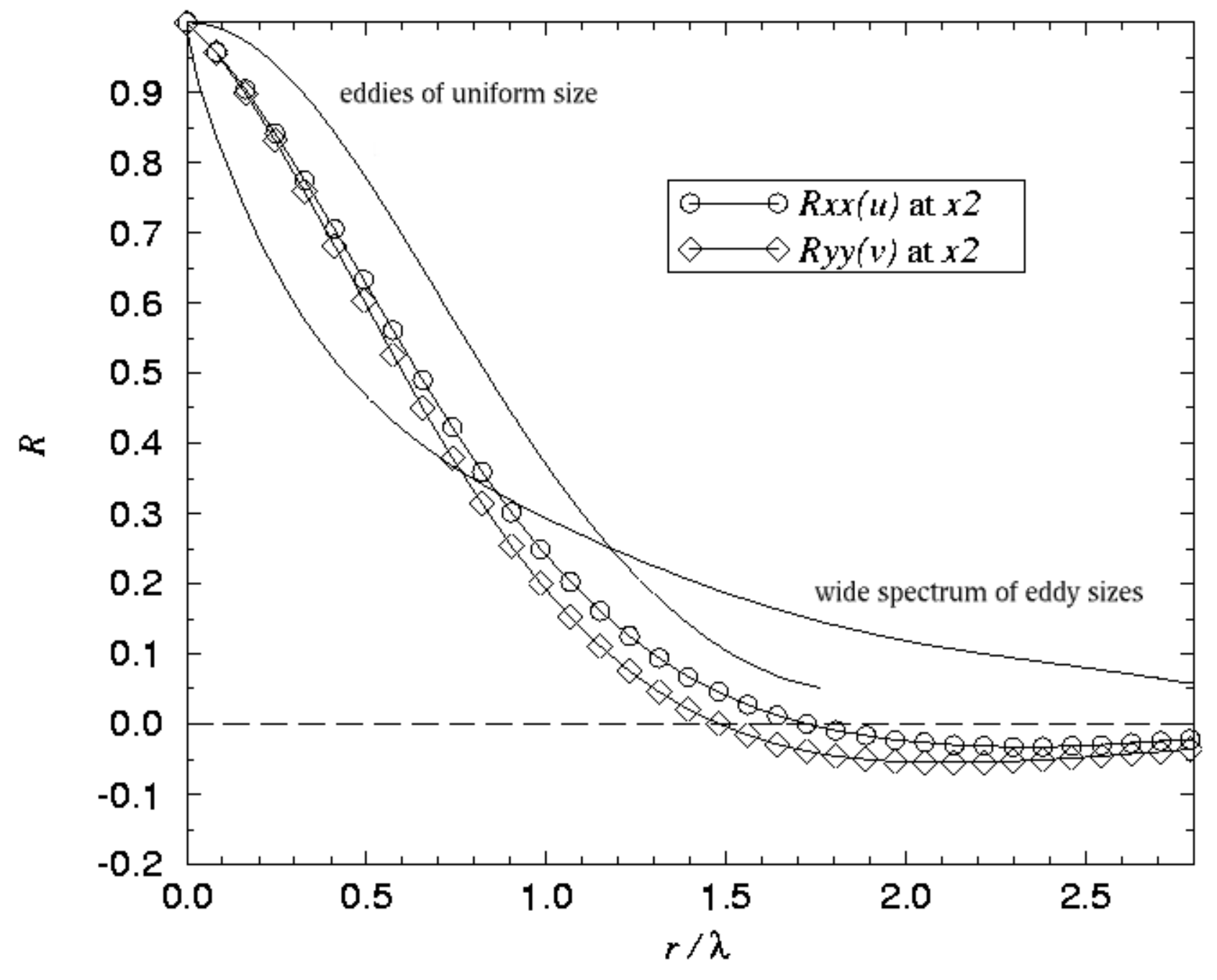}
\caption{Longitudinal autocorrelation functions of $u$ and $v$ for data set A2, $Re_\lambda = 23$ (top) and C2, $Re_\lambda = 9$ (bottom)}\label{fig:auto-long-vel-A2+C2}
\end{figure}

\begin{figure}[t!]
\centering
\includegraphics[width = 120mm, angle = 0]{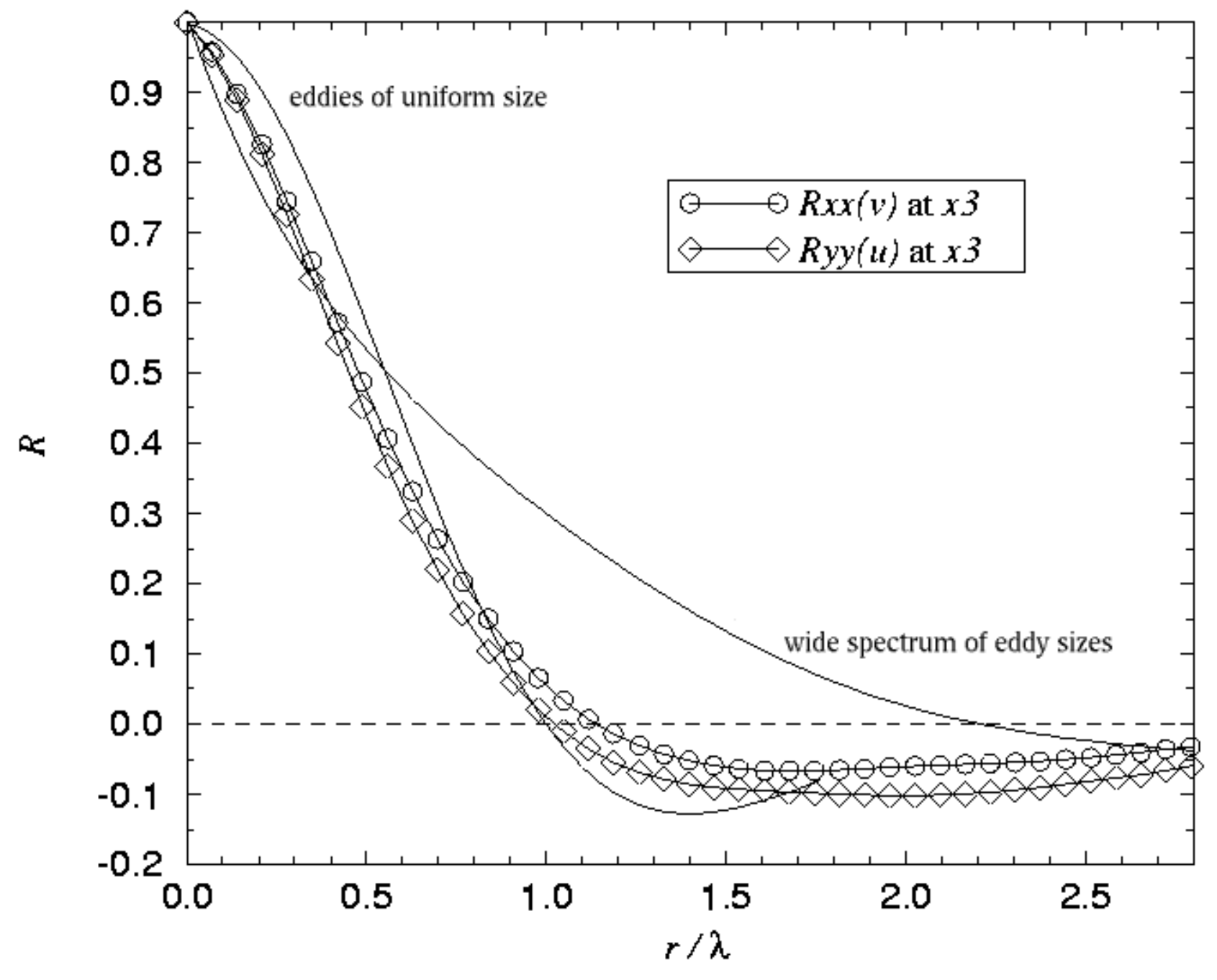} \\
\includegraphics[width = 120mm, angle = 0]{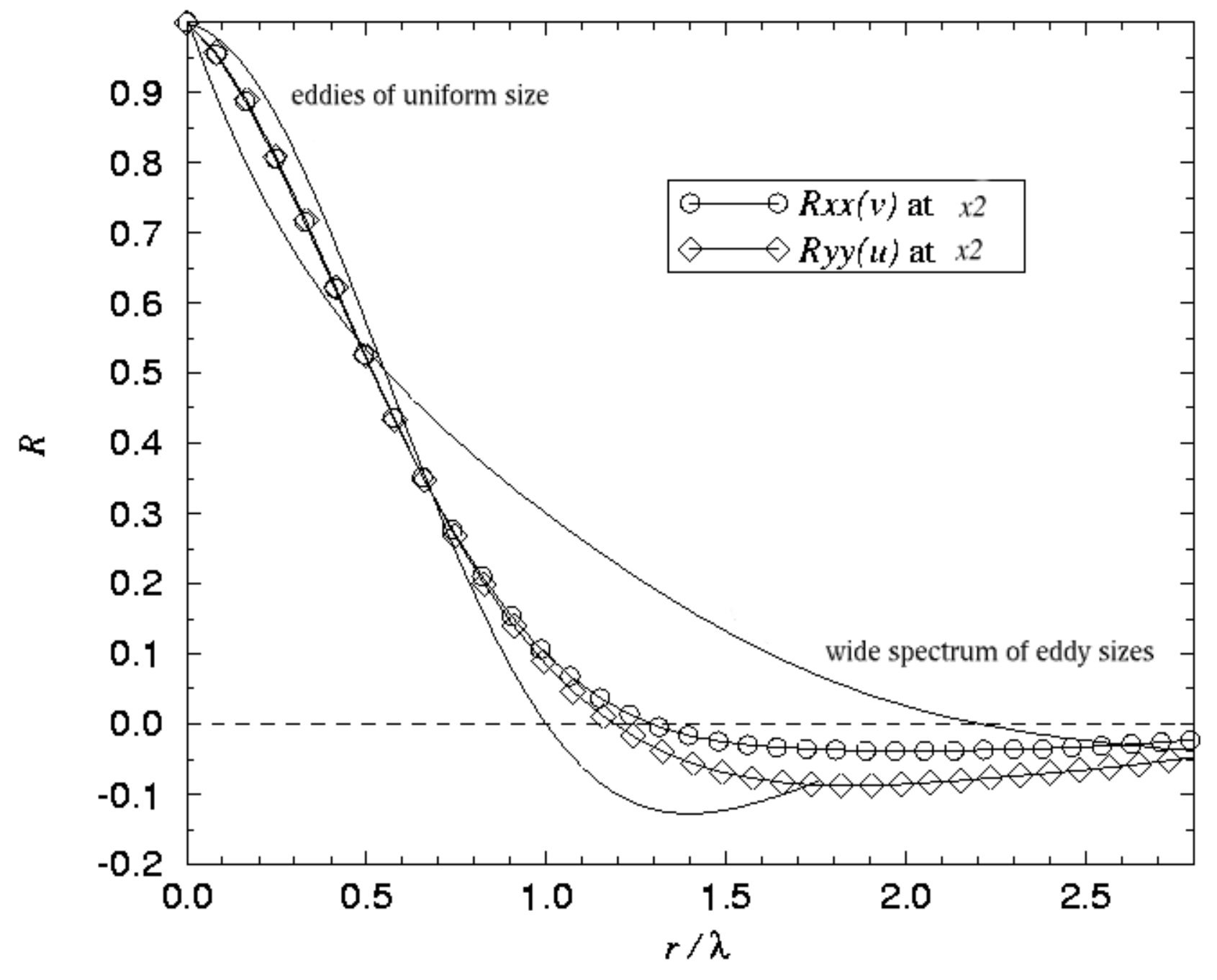} 
\caption{Transverse autocorrelation functions of $u$ and $v$ for data set A2, $Re_\lambda = 23$ (top) and C2, $Re_\lambda = 9$ (bottom)}\label{fig:auto-trans-vel-A2+C2}
\end{figure}

The longitudinal and transverse autocorrelation functions for $u$ and $v$ for data set A2, $Re_\lambda = 23$, and C2, $Re_\lambda = 9$, are shown in Figures \ref{fig:auto-long-vel-A2+C2} and \ref{fig:auto-trans-vel-A2+C2} respectively. For all cases the longitudinal autocorrelation function reaches a value of $R=0$ at approximately $r=1.5\lambda$, while the transverse autocorrelation function reaches a value of $R=0$ at approximately $r=1\lambda$.

Also shown in Figures \ref{fig:auto-long-vel-A2+C2} and \ref{fig:auto-trans-vel-A2+C2} are the autocorrelation functions found by Townsend (1976) for isotropic turbulence with uniform size structures and turbulence with a wide range of structure sizes. The experimentally determined velocity autocorrelation functions appear to more closely resemble those depicted by Townsend for uniform size structures. This differs from the results obtained in other measurements of grid turbulence. In the work of Snyder and Lumley (1971) and Comte-Bellot and Corrsin (1971), autocorrelation functions that are more consistent with a wide range of structure sizes were found. The present results suggest that larger structures are not influencing the autocorrelation to the same degree as was found in previous studies. In this study the size of each spatial velocity field was typically $3\lambda$ x $3\lambda$. If structures larger than $3\lambda$ are present then spatial averaging will remove much of their pattern. This will then produce significantly greater autocorrelation values for structures whose size is less than the imaging region, ie less than $3\lambda$.

\subsection{Velocity Gradients and Vorticity Fields}

\subsubsection{Vorticity decay}

Batchelor and Townsend (1947) showed that:
\begin{equation}
\frac{d(\frac{1}{\omega '})}{dt}=constant
\end{equation}

For grid turbulence this equation can be rewritten:
\begin{equation}
\frac{d(\frac{1}{\omega '})}{dx}=constant
\end{equation}

\begin{figure}[t!]
\centering
\includegraphics[width = 120mm, angle = 0]{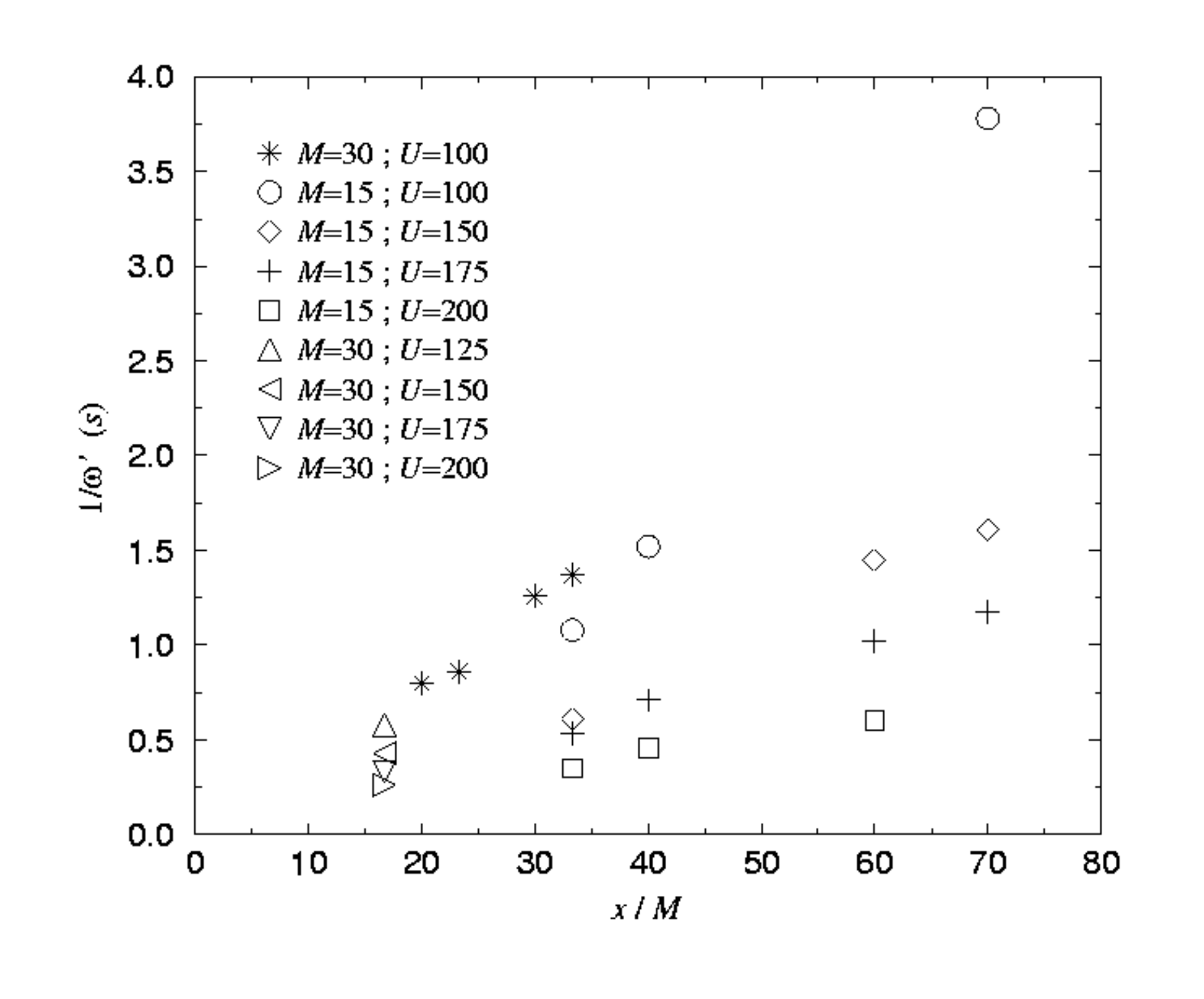}
\caption{Variation of $1/\omega '$ downstream from the grid}
\label{fig:vortdecay}
\end{figure}

In Figure \ref{fig:vortdecay} $(1/\omega ')$ is plotted against $x/M$ for the current data, and appears to show that vorticity decays in an inverse manner. This relationship was also verified experimentally by Kit et.al. (1988). 

\begin{figure}[t!]
\centering
\includegraphics[width = 120mm, angle = 0]{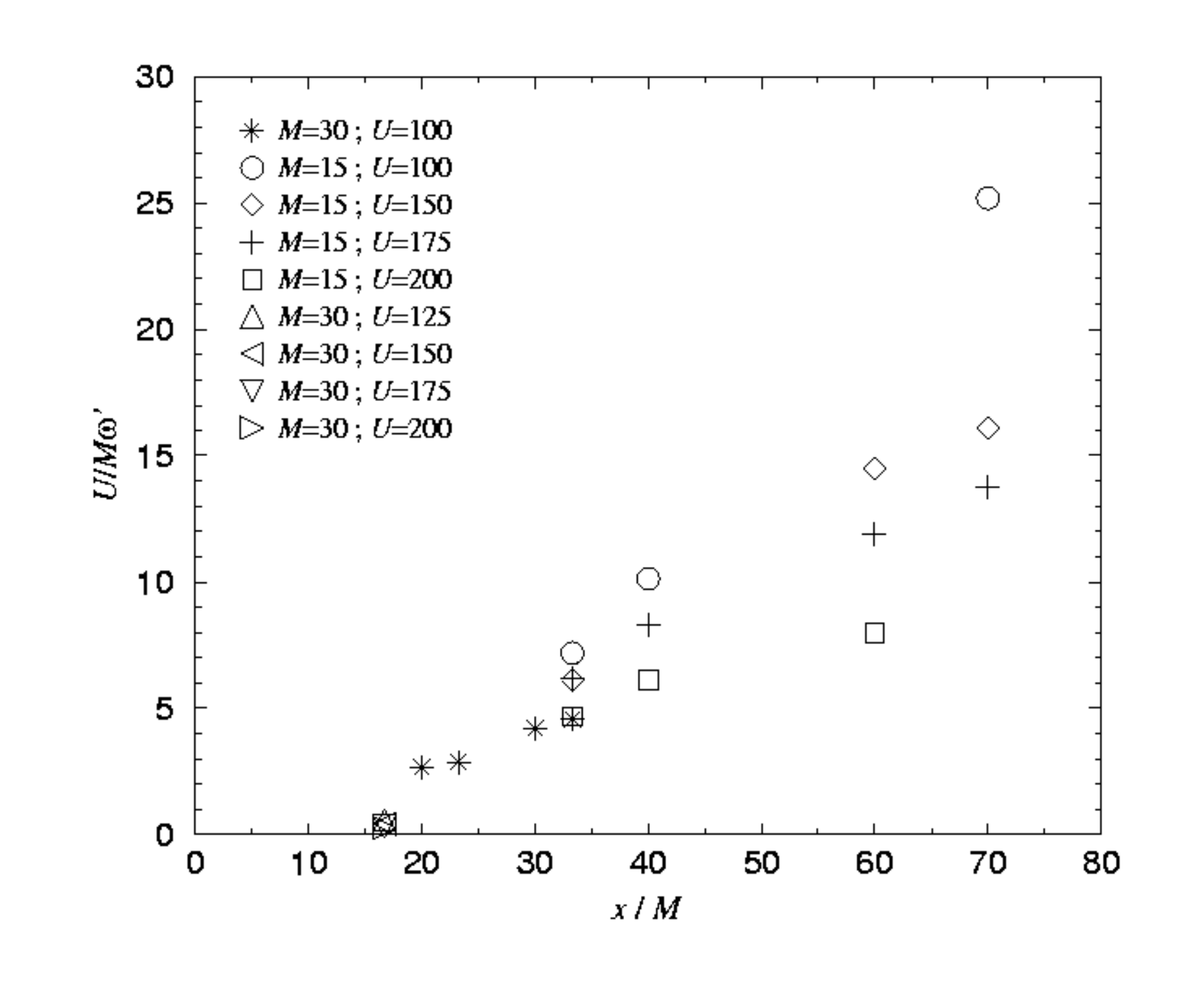}
\caption{Variation of $U/M\omega '$ downstream from the grid}
\label{fig:nondvortdecay}
\end{figure}

If the vorticity is converted to a non-dimensional variable ($U/M\omega '$), the result, plotted against $x/M$, is shown in Figure \ref{fig:nondvortdecay}. Figures \ref{fig:nondvortdecay} and \ref{fig:virtorig} are quite similar, i.e. $U/M\omega '$ as a function of $x/M$ exhibits similarity to $(U/\bf{u})^2$ as a function of $x/M$. 

\subsubsection{Probability Distribution Functions}
\begin{figure}[t!]
\centering
\includegraphics[width = 120mm, angle = 0]{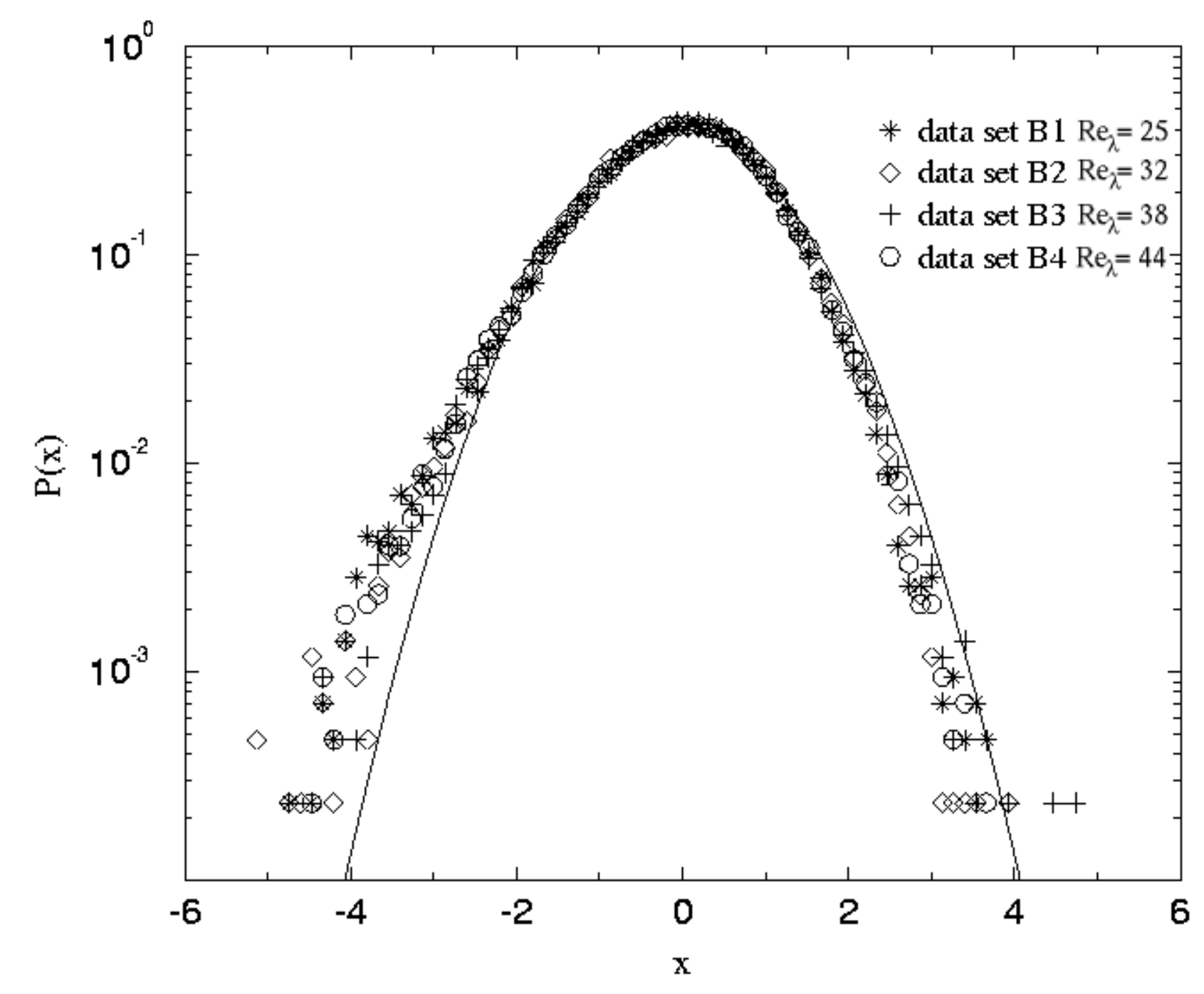}\\
\includegraphics[width = 120mm, angle = 0]{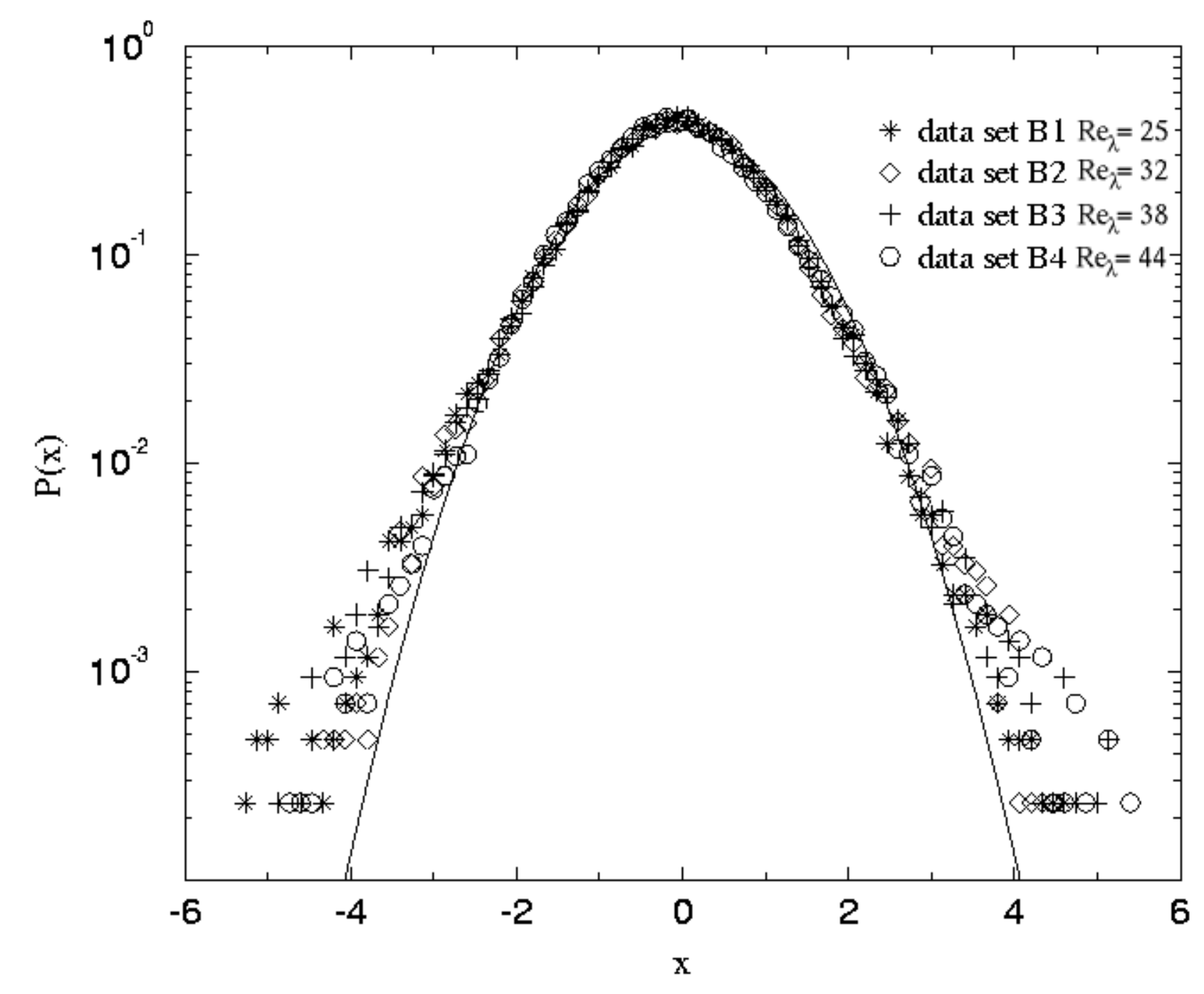} 
\caption{Probability distribution functions of longitudinal velocity gradients for data set B, normalised so that $\sigma=1$. The range of $Re_\lambda$ represented is from 25 to 44. The Gaussian distribution is shown as a solid line. $\frac{du}{dx}$ (top); $\frac{du}{dy}$ (bottom)}\label{fig:grad-pdf-B}
\end{figure}

PDFs for the velocity gradients and vorticity were calculated in the same way as those for fluctuating velocity. Figure \ref{fig:grad-pdf-B} shows the PDFs for the longitudinal and transverse gradients of the $x$-component of velocity ($u$) for data set B, $Re_\lambda$ between 25 and 44. These results are typical of the results obtained for all the experimental conditions. The longitudinal gradient ($\frac{du}{dx}$) shows a decidedly negative skew in the data, while the transverse gradient ($\frac{du}{dy}$) shows non-Gaussian tails. Calculation of skewness shows that for $\frac{du}{dx}$ skewness values are generally negative, but values between -0.4 and 0.1 are found, and kurtosis values between 3.3 and 4.3 are generally found. This compares favourably to the results of Townsend (1947) who reported a skewness of -0.38 and kurtosis of 3.5 for the PDF of $\frac{du}{dx}$. For the transverse velocity gradient ($\frac{du}{dy}$) skewness values between -0.4 and 0.4, and kurtosis values between 3.7 and 4.2 are generally found.

DNS data for isotropic turbulence indicates that the skewness of transverse velocity gradients is zero (Vincent and Meneguzzi (1991), Jimenez et.al. (1993)). Jimenez et.al. (1993) observed that the transverse velocity gradients are more symmetrical than the longitudinal velocity gradients. This is also apparent in the experimental data obtained in this study. Jimenez et.al (1993) also observed an increase in magnitude of the kurtosis with $Re_{\lambda}$ for both the longitudinal and transverse velocity gradients. In this experimental study, a small increase in kurtosis for $dv/dy$ was found at increasing values of  $Re_{\lambda}$, but this result was not duplicated for the other velocity gradients.

\begin{figure}[t!]
\centering
\includegraphics[width = 120mm, angle = 0]{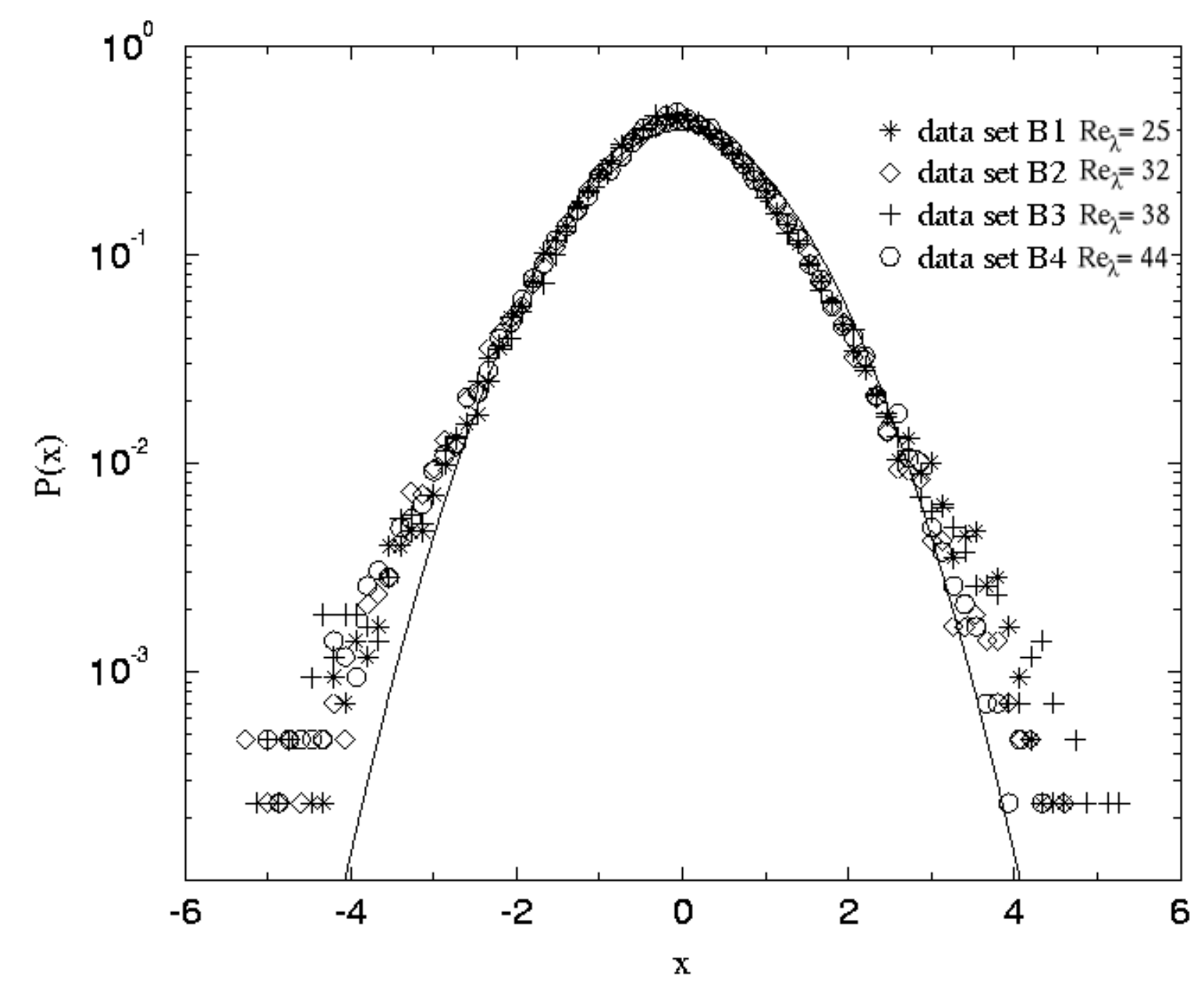}\\
\caption{Probability distribution functions of vorticity for data set B, normalised so that $\sigma=1$. The Gaussian distribution is shown as a solid line.}\label{fig:vort-pdf-B}
\end{figure}

Figure \ref{fig:vort-pdf-B} shows the PDF of vorticity for data set B, $Re_\lambda$ between 25 and 44. This PDF shows a slight negative skew and the presence of non-Gaussian tails in the distribution. The skewness values are quite small and range between -0.2 an 0.1, and the kurtosis values are between 3.7 and 4.3. 

\subsubsection{Autocorrelation Functions}
\begin{figure}[t!]
\centering
\includegraphics[width = 120mm, angle = 0]{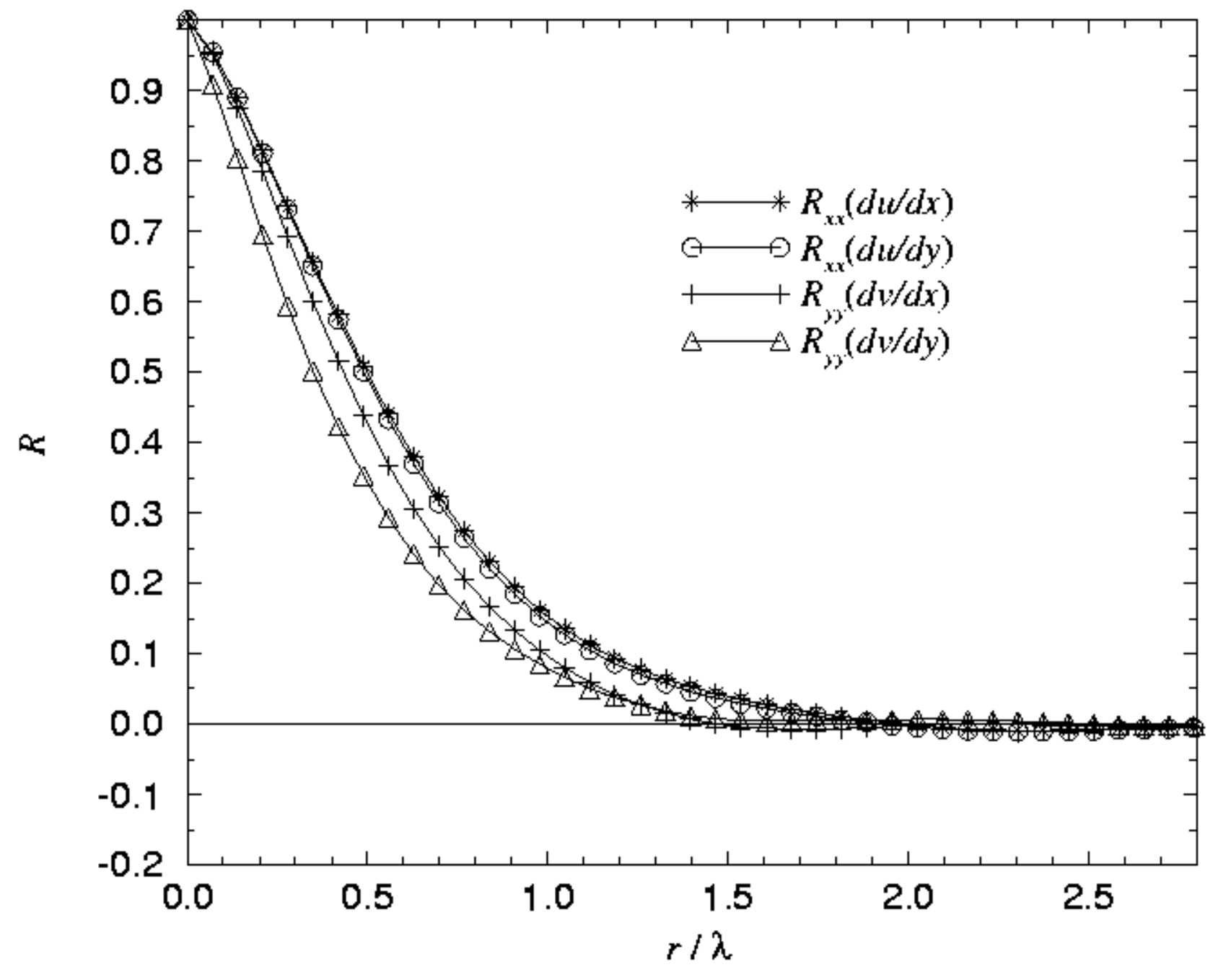}\\
\includegraphics[width = 120mm, angle = 0]{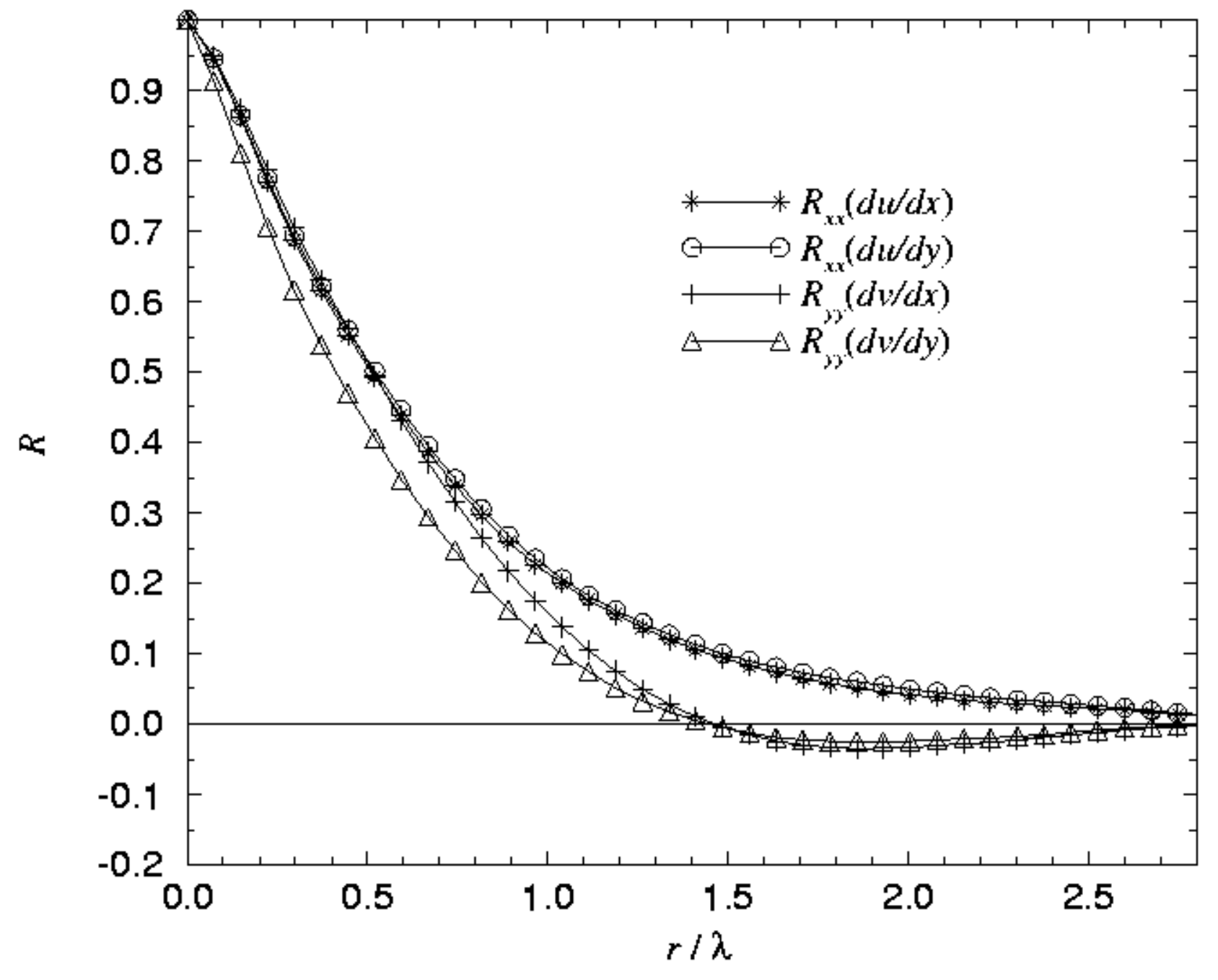} 
\caption{Longitudinal autocorrelation functions, direction defined with respect to velocity, of velocity gradients for data set A2, $Re_\lambda = 23$ (top) and A3, $Re_\lambda = 16$ (bottom).}
\label{fig:grad-long-auto-A2+A3}
\end{figure}

\begin{figure}[t!]
\centering
\includegraphics[width = 120mm, angle = 0]{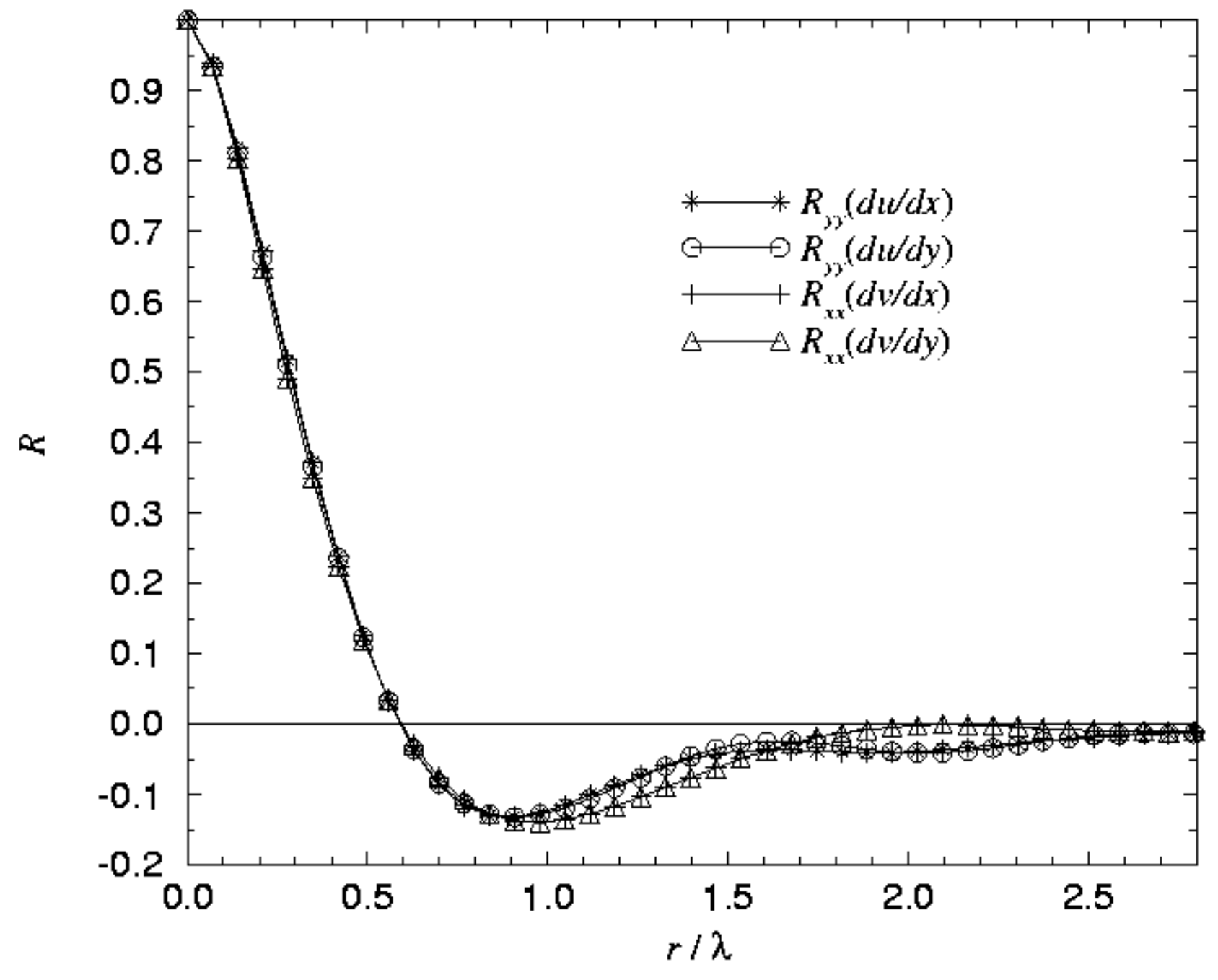}\\
\includegraphics[width = 120mm, angle = 0]{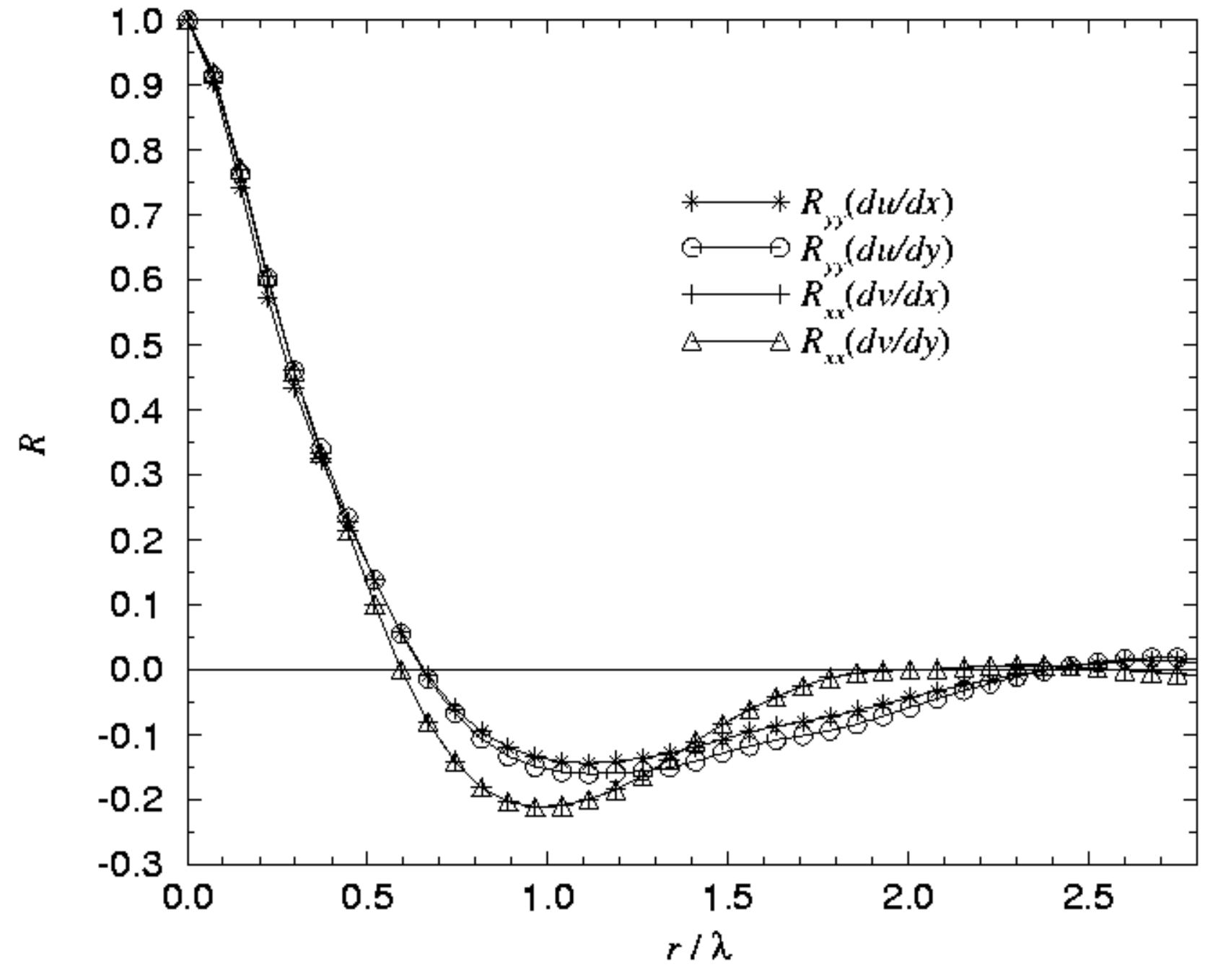} 
\caption{Transverse autocorrelation functions, direction defined with respect to velocity, of velocity gradients for data set A2, $Re_\lambda = 23$ (top) and A3, $Re_\lambda = 16$ (bottom).}
\label{fig:grad-trans-auto-A2+A3}
\end{figure}

Velocity gradient and vorticity autocorrelations are calculated in the same way as those for velocity. The longitudinal autocorrelation functions for the velocity gradient for data sets A2, $Re_\lambda = 23$, and A3, $Re_\lambda = 16$, are shown in Figure \ref{fig:grad-long-auto-A2+A3}, and the corresponding transverse autocorrelation functions are shown in Figure \ref{fig:grad-trans-auto-A2+A3}. The direction of the autocorrelation function is defined with respect to the velocity direction. The transverse autocorrelations of the velocity gradients are observed to have a value of $R=0$ at approximately $r=0.5\lambda$. The Lagrangian autocorrelations given by Yeung and Pope (1989) for acceleration components taken from DNS data of isotropic turbulence have a similar shape to the longitudinal autocorrelations of the velocity gradients shown here. The shape of the transverse autocorrelations of the velocity gradients also conforms to that suggested by Tennekes and Lumley (1972) for the temporal autocorrelation function of the derivative of a velocity component.  

\begin{figure}[t!]
\centering
\includegraphics[width = 120mm, angle = 0]{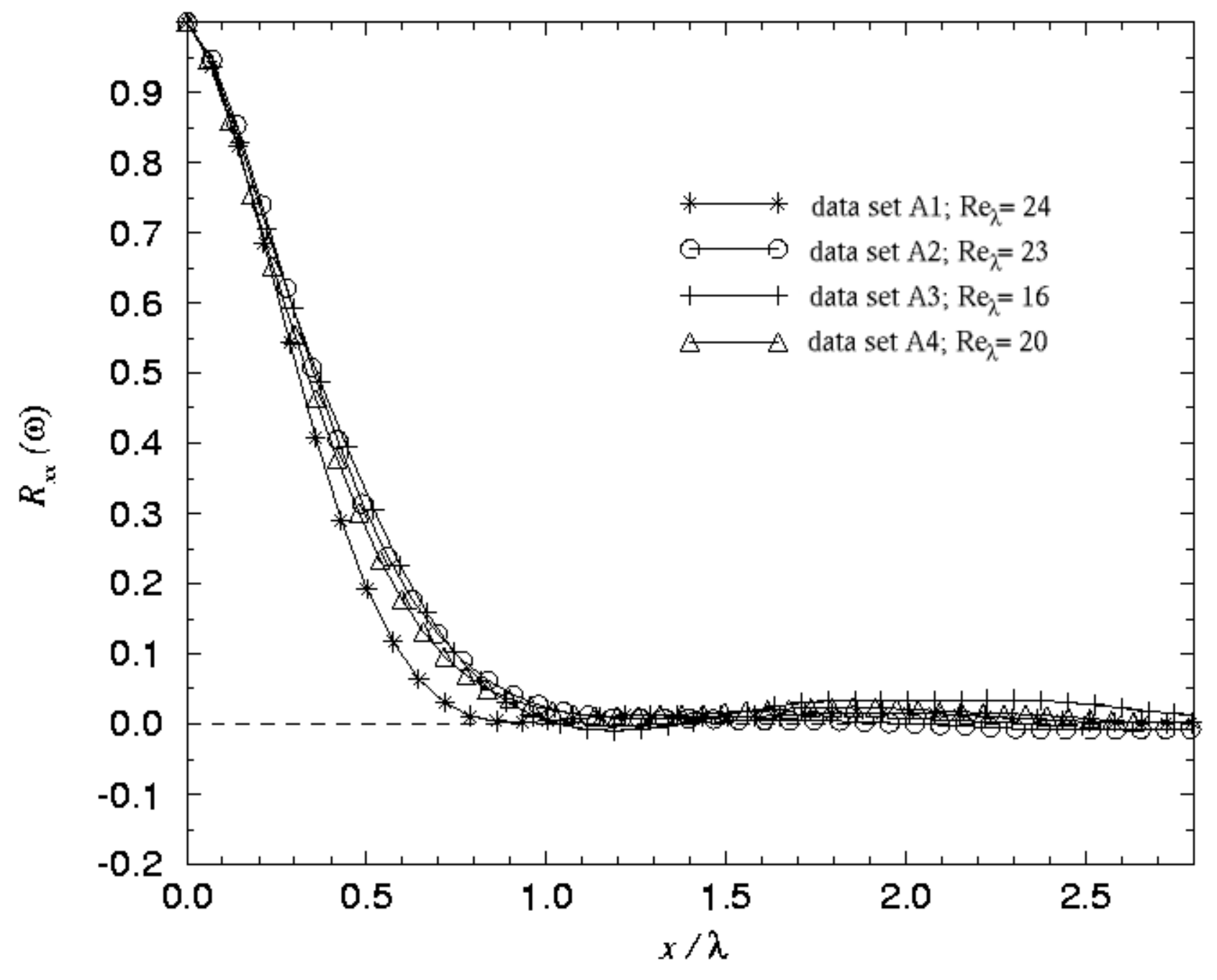}\\
\caption{Streamwise autocorrelation of $\omega$ for data set A. The range of $Re_\lambda$ represented is from 16 to 24.}
\label{fig:vort-stream-auto-A}
\end{figure}

\begin{figure}[t!]
\centering
\includegraphics[width = 120mm, angle = 0]{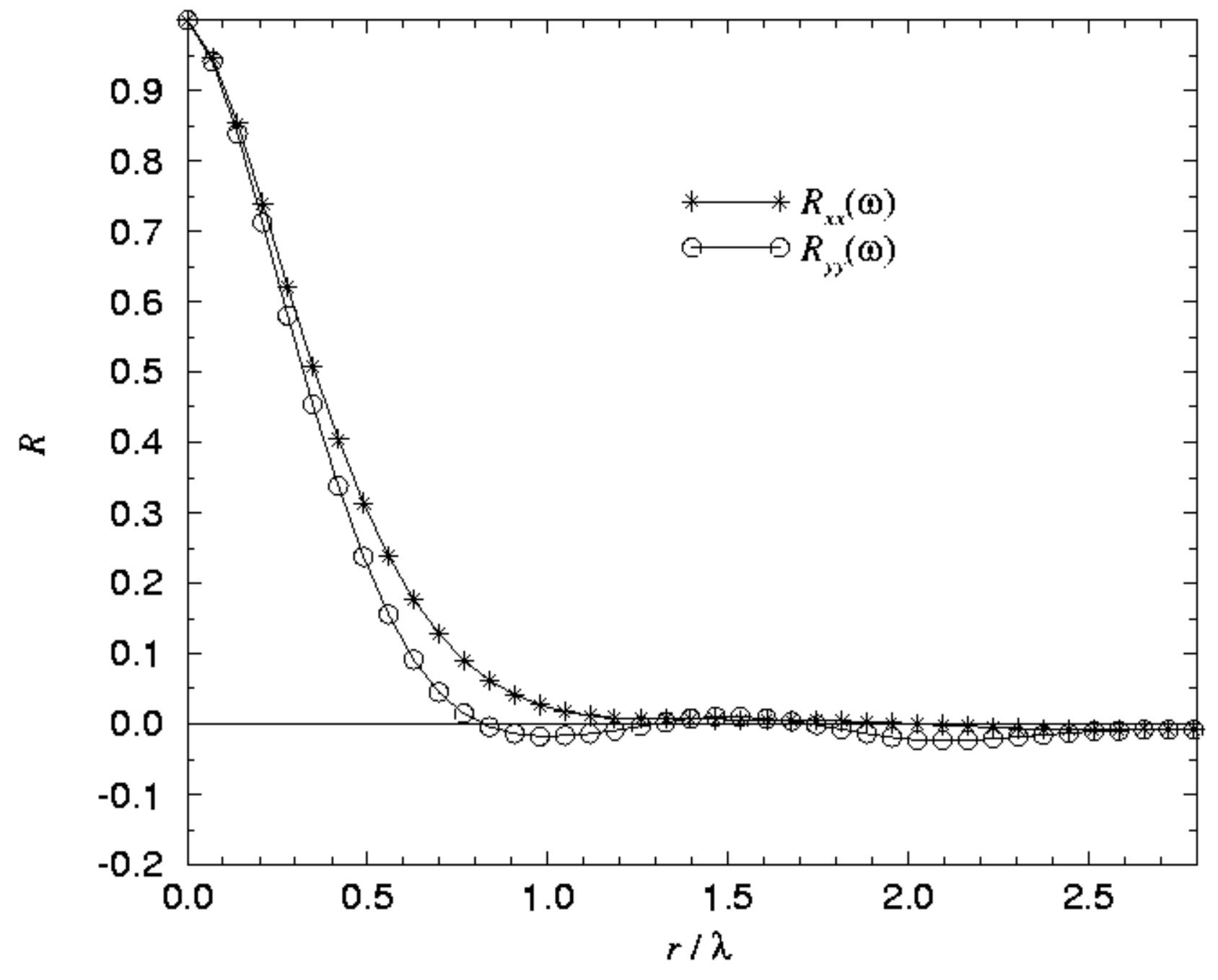}\\
\includegraphics[width = 120mm, angle = 0]{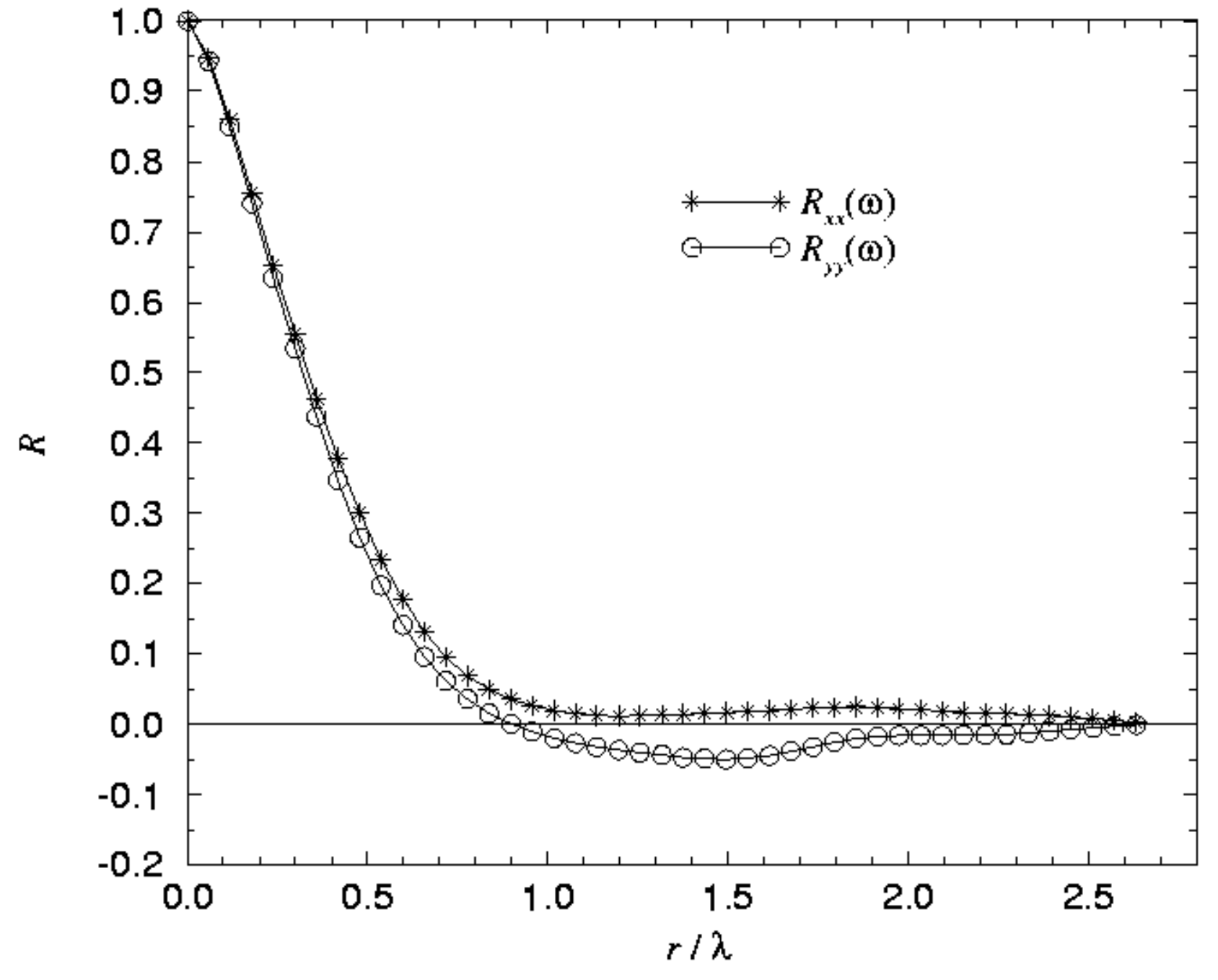} 
\caption{Vorticity autocorrelation functions for data set A2, $Re_\lambda = 23$ (top) and A4, $Re_\lambda = 20$ (bottom).}
\label{fig:vort-auto-A2+A4}
\end{figure}

The streamwise autocorrelations of $\omega_z$, $R_{xx}(\omega)$ for data set A is shown in Figure \ref{fig:vort-stream-auto-A}, and Figure \ref{fig:vort-auto-A2+A4} shows the streamwise and cross-stream autocorrelations of the vorticity for two data sets. Note that these autocorrelations are in the transverse direction with respect to the vorticity.

\section{Discussion}
The PDFs generated from the PIV data for velocity follow closely those obtained from past experimental investigations and DNS studies. However, the autocorrelation functions obtained for the velocity differ from past results. The reason for the discrepancy is conjectured to be the size of the spatial domain. This was alluded to earlier when comparing the autocorrelation functions obtained for velocity to those obtained by Townsend for turbulence with a wide range of structures, and turbulence with uniform size structures.

Data obtained from DNS of three dimensional homogeneous, isotropic turbulence was used to investigate the effect of spatial domain size on the autocorrelation functions. All of the DNS results presented are from a simulation with $Re_\lambda = 41$. Details of the DNS code can be found in O'Neill and Soria (2004). The size of the simulation grid was $2\pi^3$ and $\lambda = 0.59$, giving a simulation volume of $10.6\lambda$.

\begin{figure}[t!]
\centering
\includegraphics[width = 120mm, angle = 0]{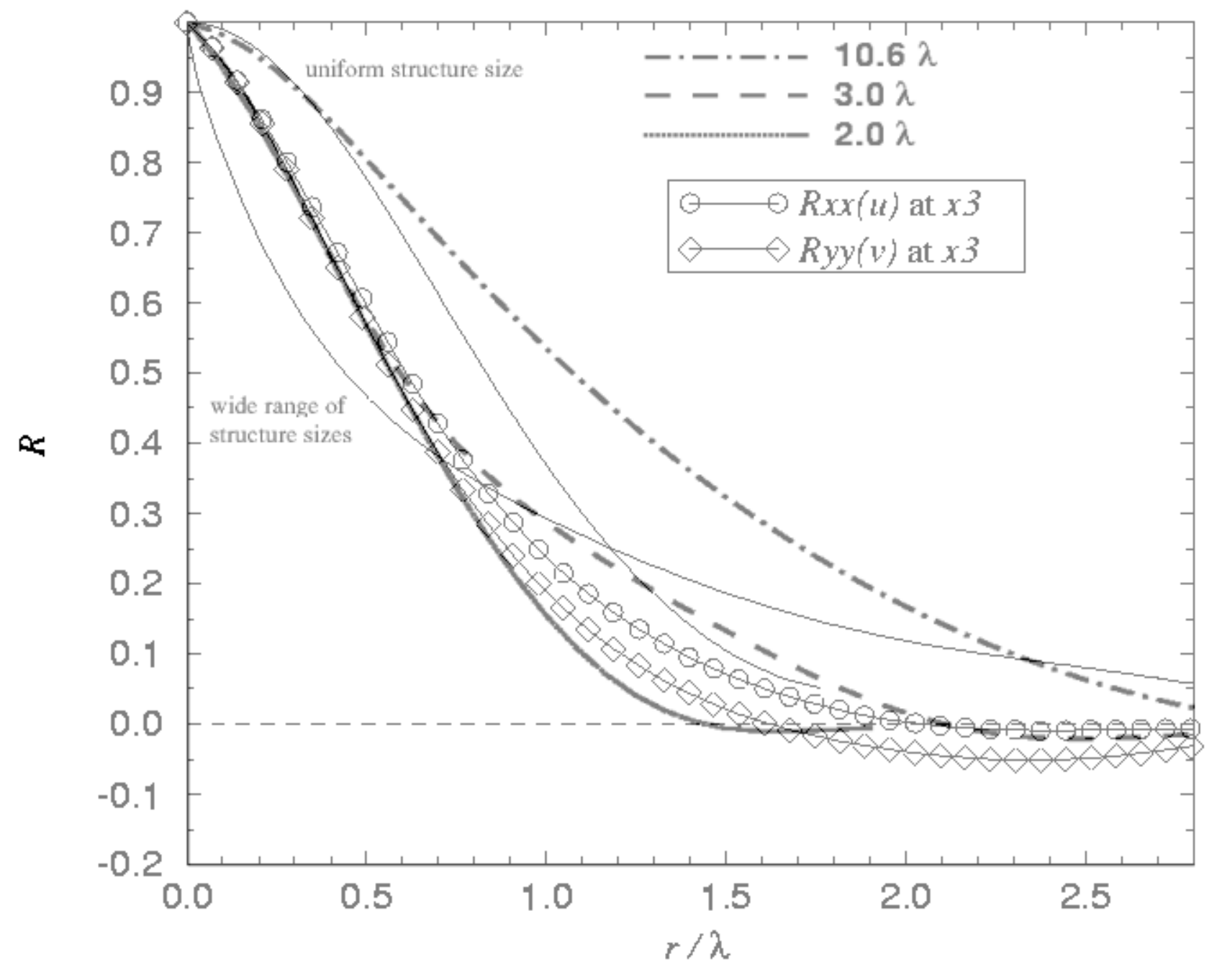}
\caption{Longitudinal velocity autocorrelations obtained from DNS data by progressively restricting the spatial domain ($Re_\lambda = 41$) compared to experimental data obtained using a spatial domain of $3\lambda$ (data set A2; $Re_\lambda = 23$)}
\label{fig:auto-long-vel-exp-dns}
\end{figure}

\begin{figure}[t!]
\centering
\includegraphics[width = 120mm, angle = 0]{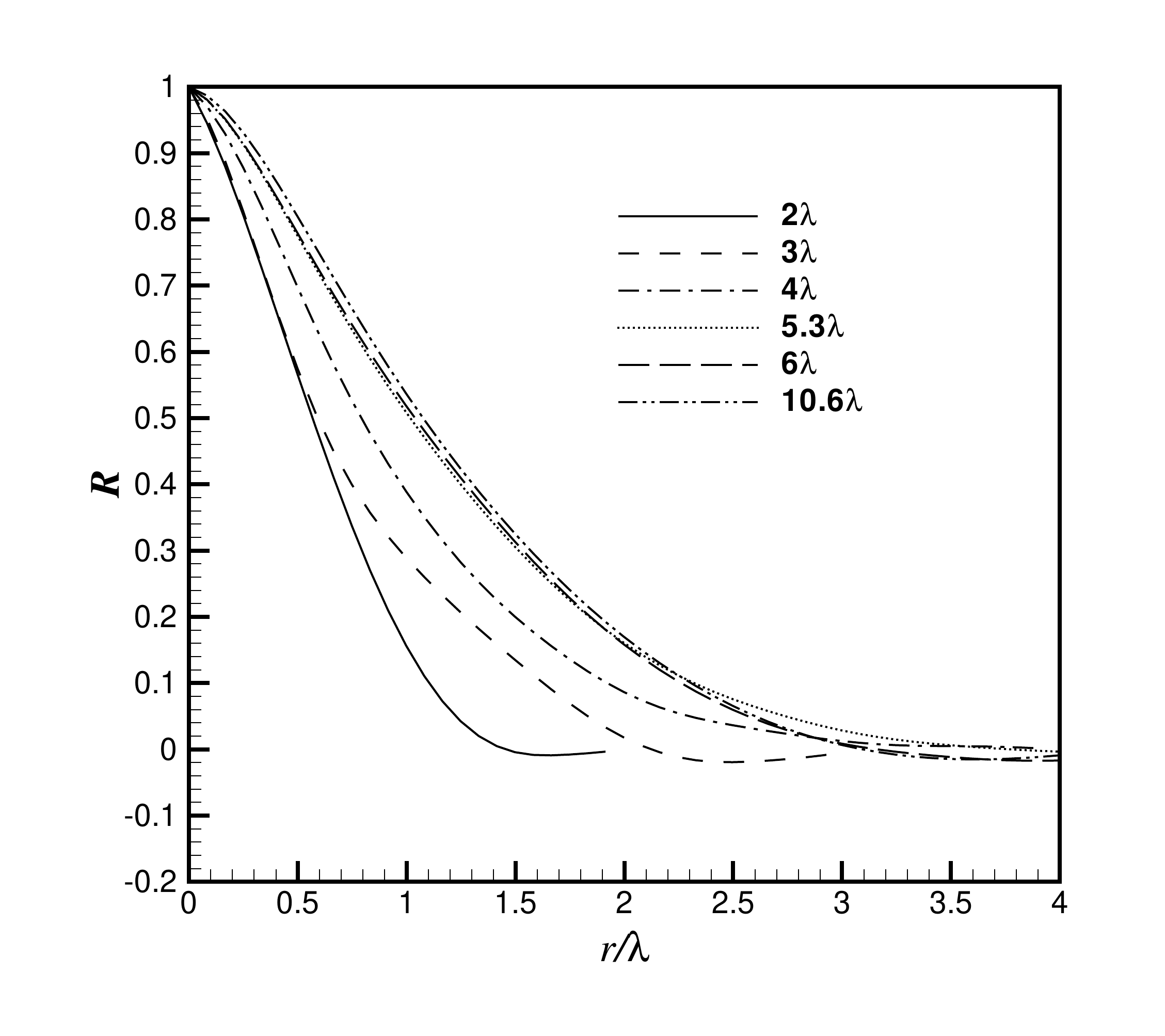}
\caption{Longitudinal velocity autocorrelations obtained from DNS data and progressively restricting the spatial domain. $Re_\lambda = 41$.}\label{fig:auto-long-vel-comp}
\end{figure}

For the same velocity field calculated on a $128^3$ mesh, the autocorrelation functions were determined for a range of spatial restrictions on the data. As the turbulence is isotropic, velocity data in each of the three dimensions can be used in the determination of the autocorrelation functions. The results obtained for the longitudinal velocity autocorrelation are shown in Figure \ref{fig:auto-long-vel-comp} and clearly show significant changes in the autocorrelation function when the spatial domain is restricted to $4\lambda$ or less. In Figure \ref{fig:auto-long-vel-exp-dns} some of these results are compared to the experimentally obtained data, in particular the DNS result for a spatial domain of $3\lambda$, which is equivalent to the experimental spatial domain. It appears that  the restricted spatial domain in the experiment accounts for much of the deviation of the PIV results from previous experimental and numerical velocity data where the spatial domain is not so restricted.

\begin{figure}[t!]
\centering
\includegraphics[width = 120mm, angle = 0]{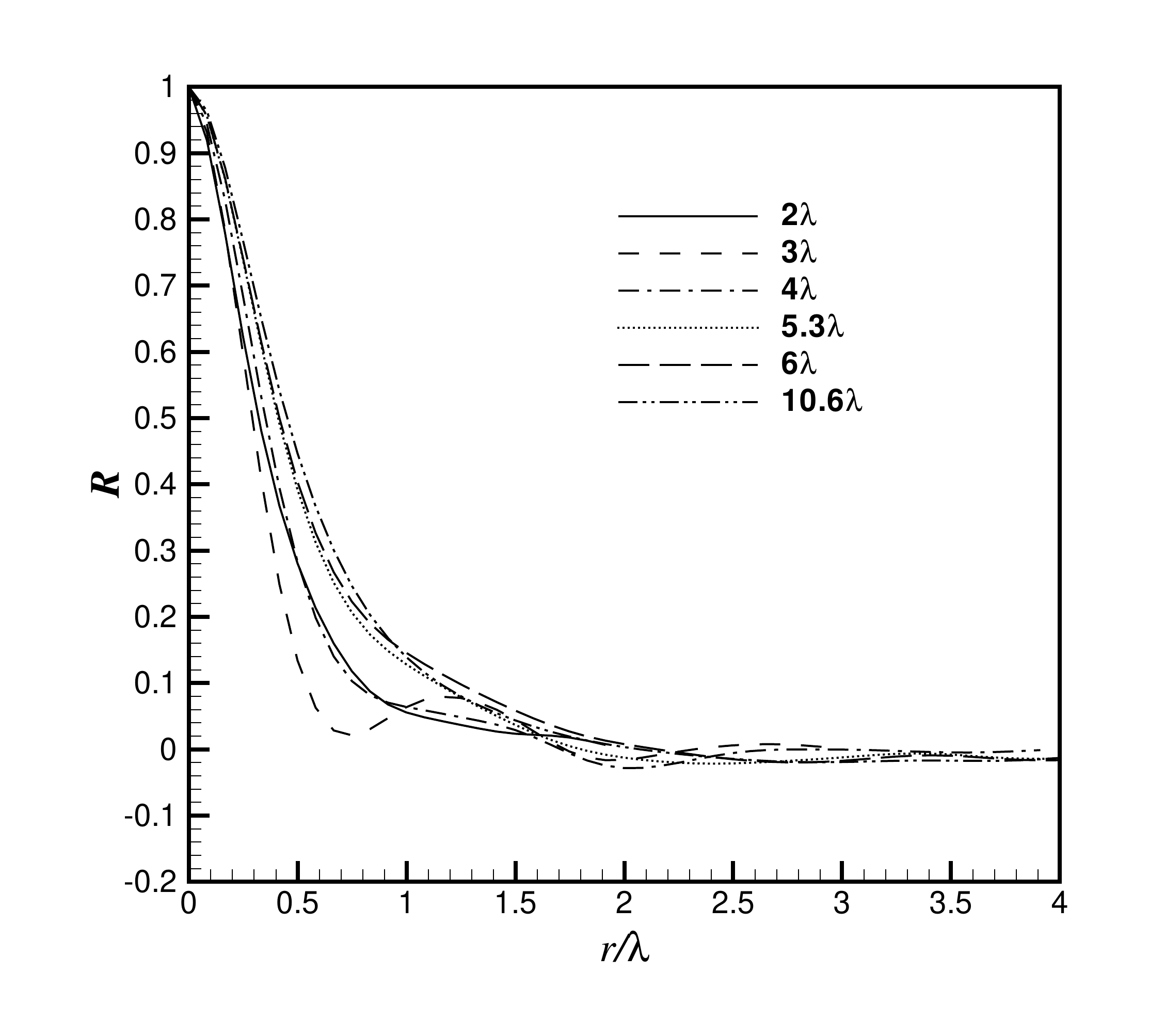}
\caption{Longitudinal velocity gradient autocorrelation functions obtained from DNS data and progressively restricting the spatial domain. $Re_\lambda = 41$}
\label{fig:auto-long-grad-comp}
\end{figure}

\begin{figure}[t!]
\centering
\includegraphics[width = 120mm, angle = 0]{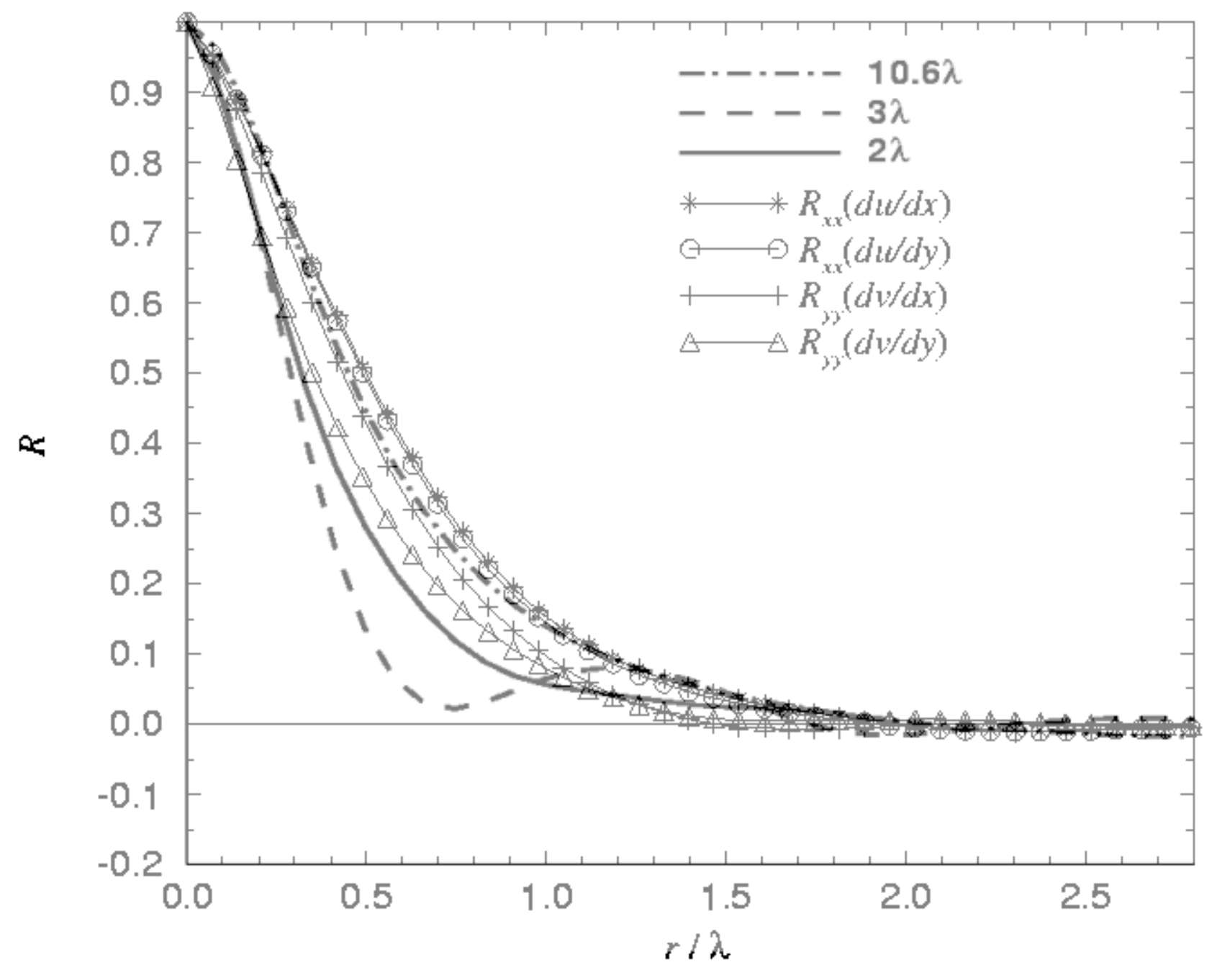}
\caption{Longitudinal velocity gradient autocorrelations obtained from DNS data by progressively restricting the spatial domain ($Re_\lambda = 41$) compared to experimental data obtained using a spatial domain of $3\lambda$ (data set A2; $Re_\lambda = 23$)}
\label{fig:auto-long-grad-exp-dns}
\end{figure}

From Figure \ref{fig:auto-long-vel-comp} it appears that restricting the spatial domain does not substantially change the shape of the autocorrelation function, but compresses that shape into a smaller linear distance. However it was found from the analysis of the DNS data that this is not a consistent result. The longitudinal autocorrelation function obtained for the velocity gradient data and restricting the spatial domain is shown in Figure \ref{fig:auto-long-grad-comp}. In this case the autocorrelations found for a spatial domain of 2$\lambda$ and 4$\lambda$ are quite similar, while that for a spatial domain of 3$\lambda$ is substantially different. Comparison between the DNS and PIV data is shown in Figure \ref{fig:auto-long-grad-exp-dns}. Here the DNS and PIV data seem to match closely, even when the spatial domain for the DNS is 10.6$\lambda$ and the spatial domain for the PIV is 3$\lambda$. 

\begin{figure}[t!]
\centering
\includegraphics[width = 120mm, angle = 0]{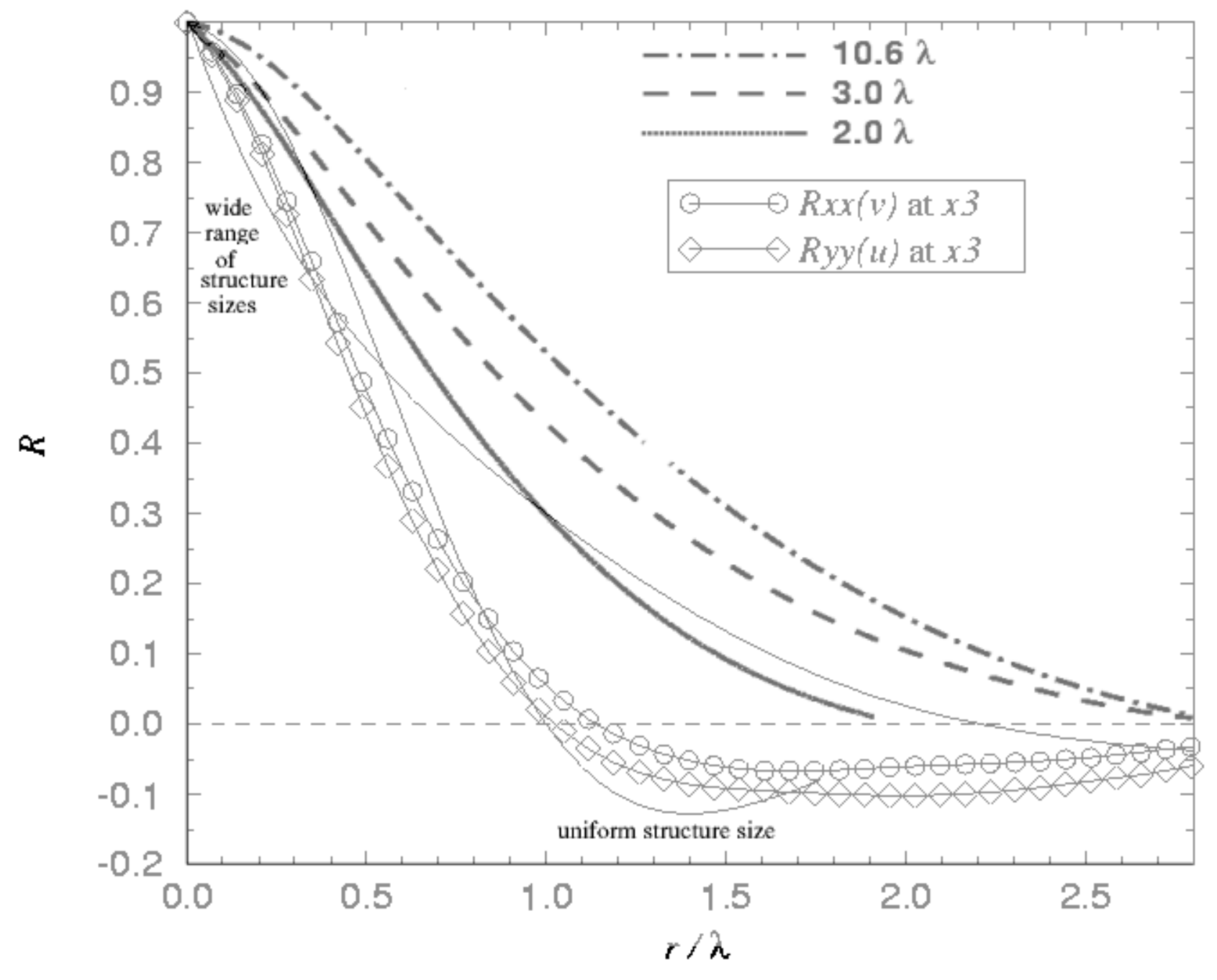}\\
\includegraphics[width = 120mm, angle = 0]{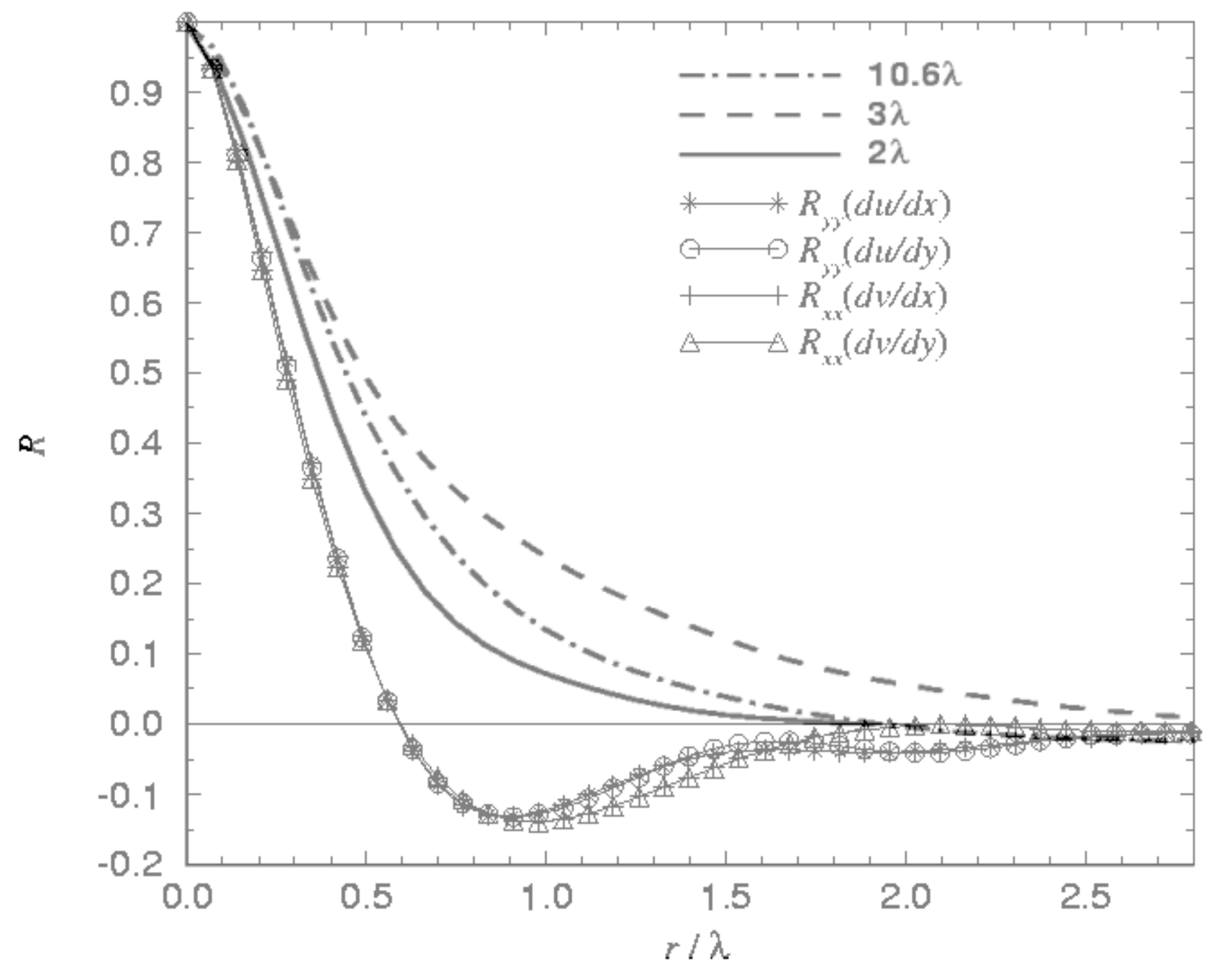}
\caption{Transverse autocorrelation functions for velocity (top) and  velocity gradient (bottom) obtained from DNS data by progressively restricting the spatial domain ($Re_\lambda = 41$) compared to experimental data obtained using a spatial domain of $3\lambda$ (data set A2; $Re_\lambda = 23$)}
\label{fig:auto-trans-vel+grad}
\end{figure}

The synchronicity between the PIV and DNS data seen for the longitudinal autocorrelation functions of velocity and velocity gradient is not replicated in the transverse autocorrelation functions of velocity and velocity gradient. Comparison of DNS data to PIV data for these two cases is shown in Figure \ref{fig:auto-trans-vel+grad}, and it can be seen that none of the autocorrelations found from the DNS data are close to the PIV data. It is also noted that for the transverse velocity gradient  the integral length, calculated by integrating the autocorrelation function, appears to be larger for a spatial domain of $3\lambda$ than for a spatial domain of $10.6\lambda$. It appears that there is no simple rule to describe the effect of restricting the spatial domain on the autocorrelation function and the integral length.

\begin{figure}[t!]
\centering
\includegraphics[width = 120mm, angle = 0]{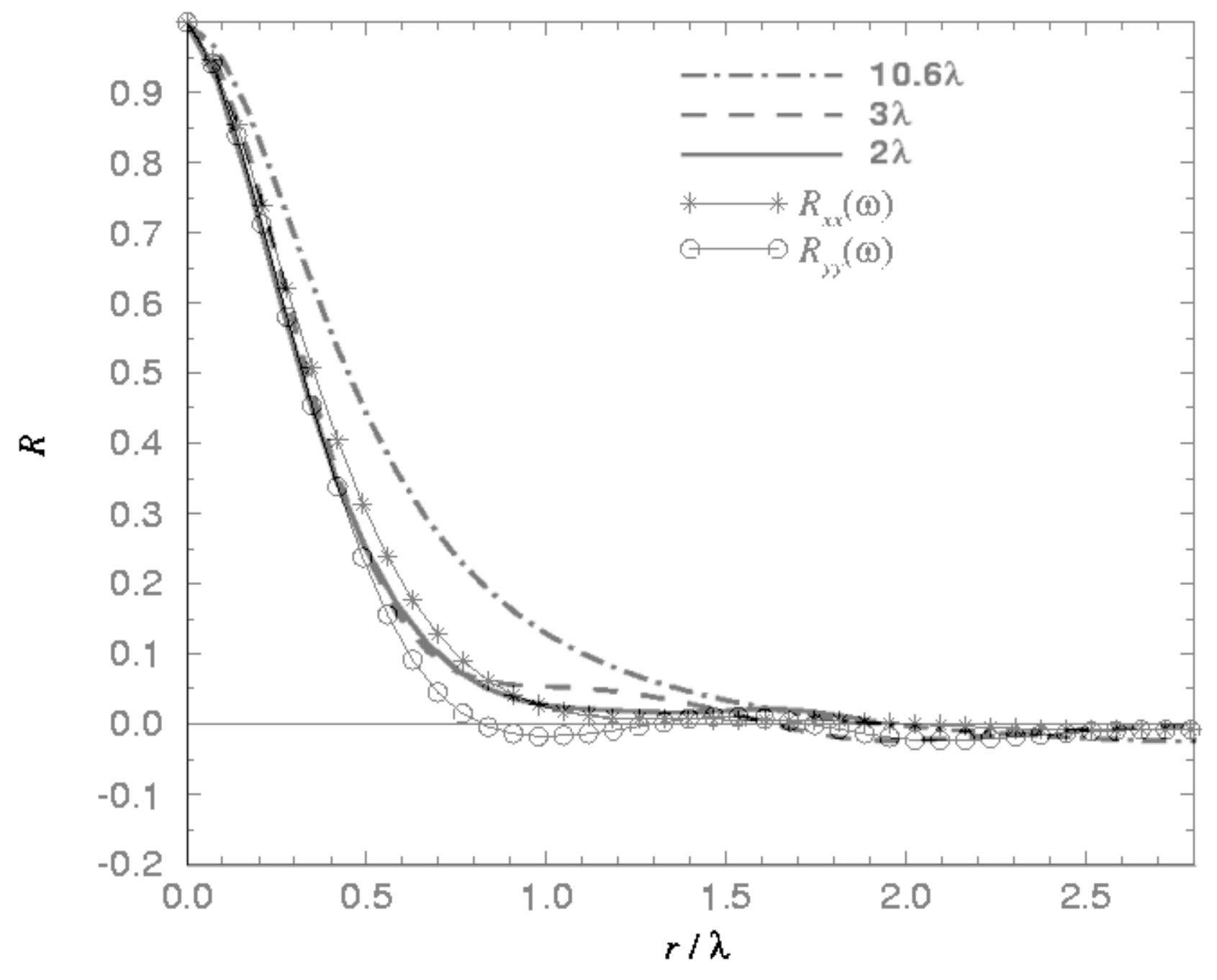}
\caption{Autocorrelation functions for vorticity obtained from DNS data by progressively restricting the spatial domain ($Re_\lambda = 41$) compared to experimental data obtained using a spatial domain of $3\lambda$ (data set A2; $Re_\lambda = 23$)}
\label{fig:auto-vort-exp-dns}
\end{figure}

The correspondance between the PIV and DNS data for the transverse autocorrelation functions of vorticity is shown in Figure \ref{fig:auto-vort-exp-dns}. As with the longitudinal autocorrelations for velocity and velocity gradient there is an excellent correspondance between the PIV and DNS data when the spatial domain is restricted to $3\lambda$.

\begin{figure}[t!]
\centering
\includegraphics[width = 80mm, angle = 0]{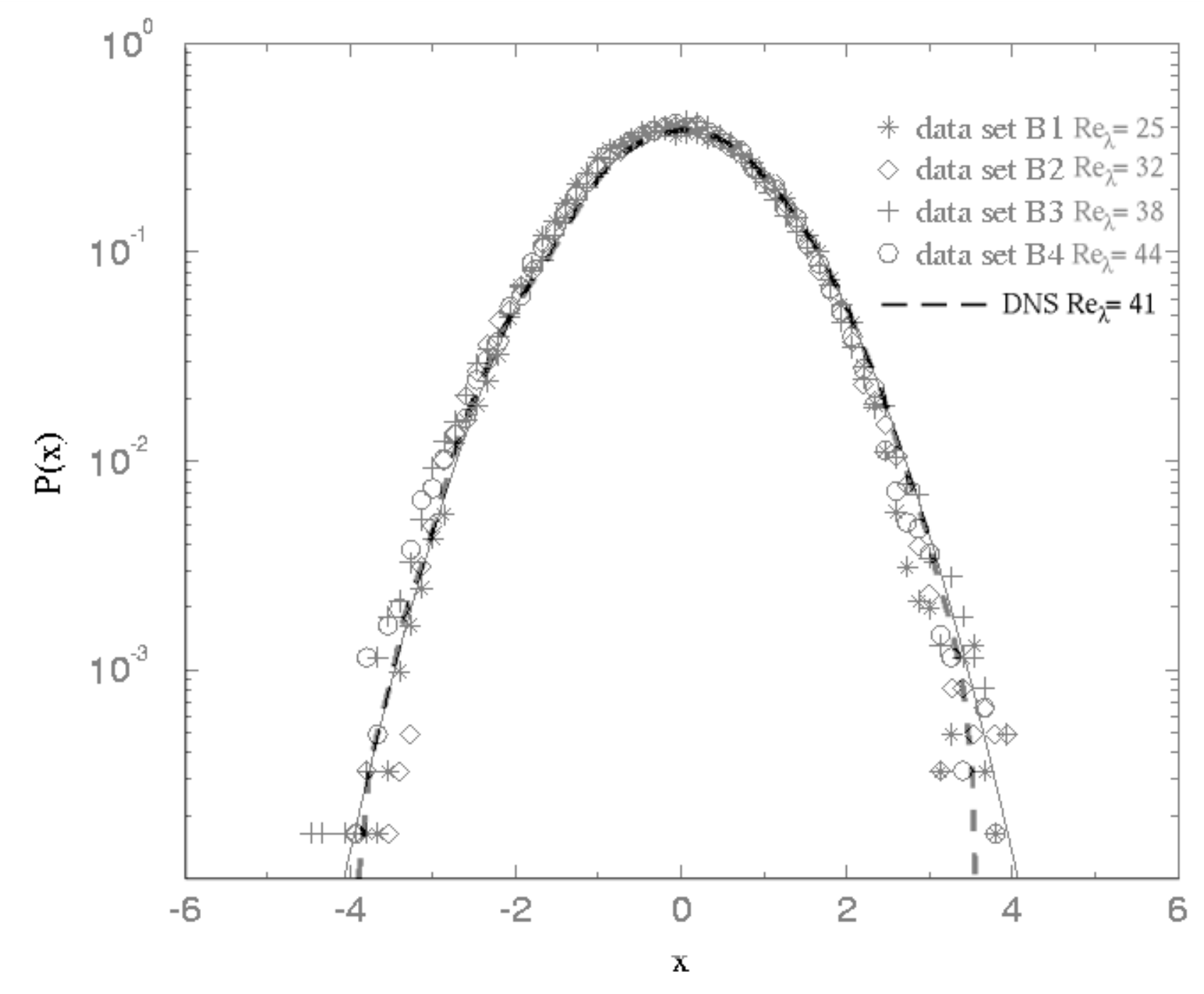}
\includegraphics[width = 80mm, angle = 0]{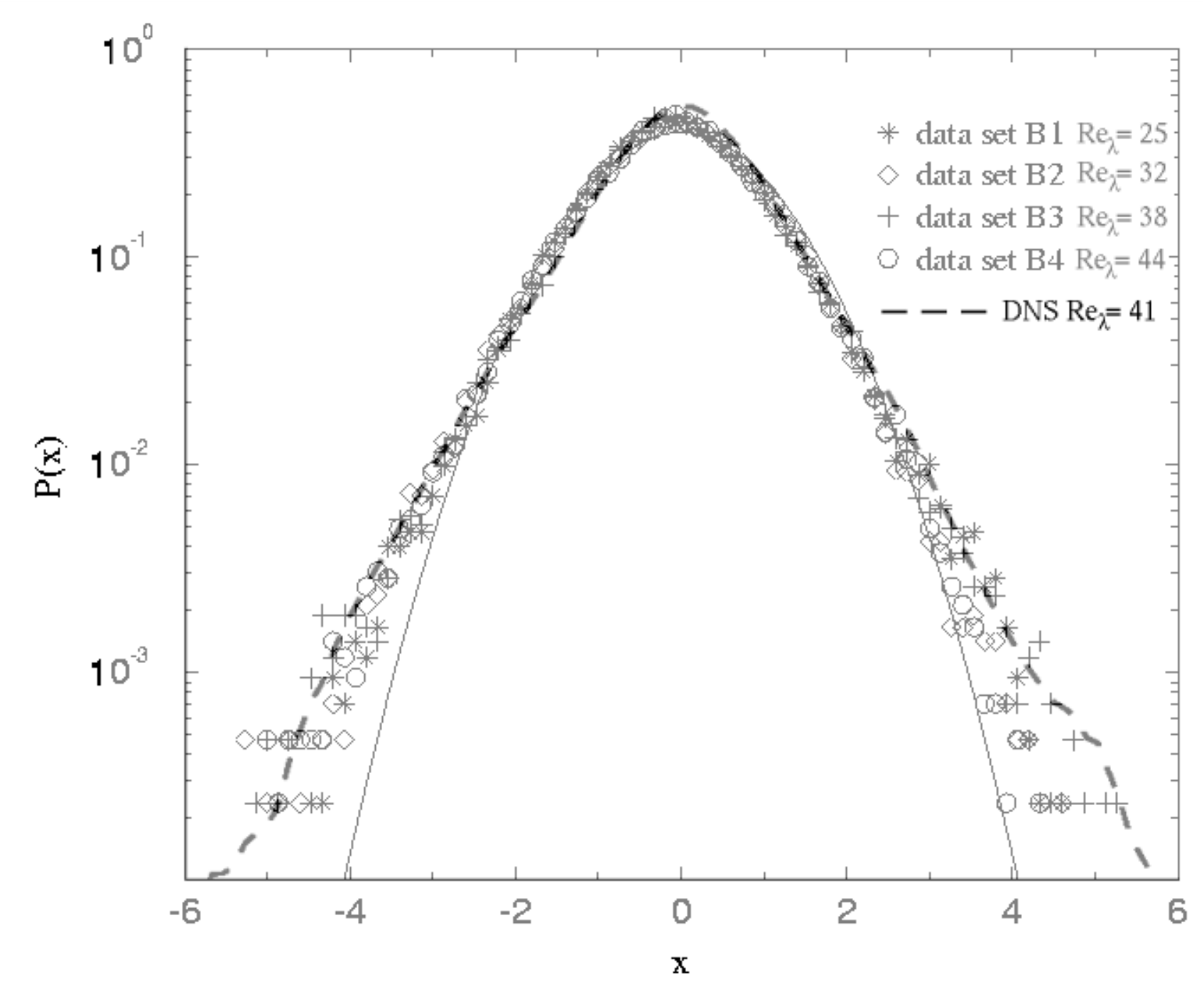}\\
\includegraphics[width = 80mm, angle = 0]{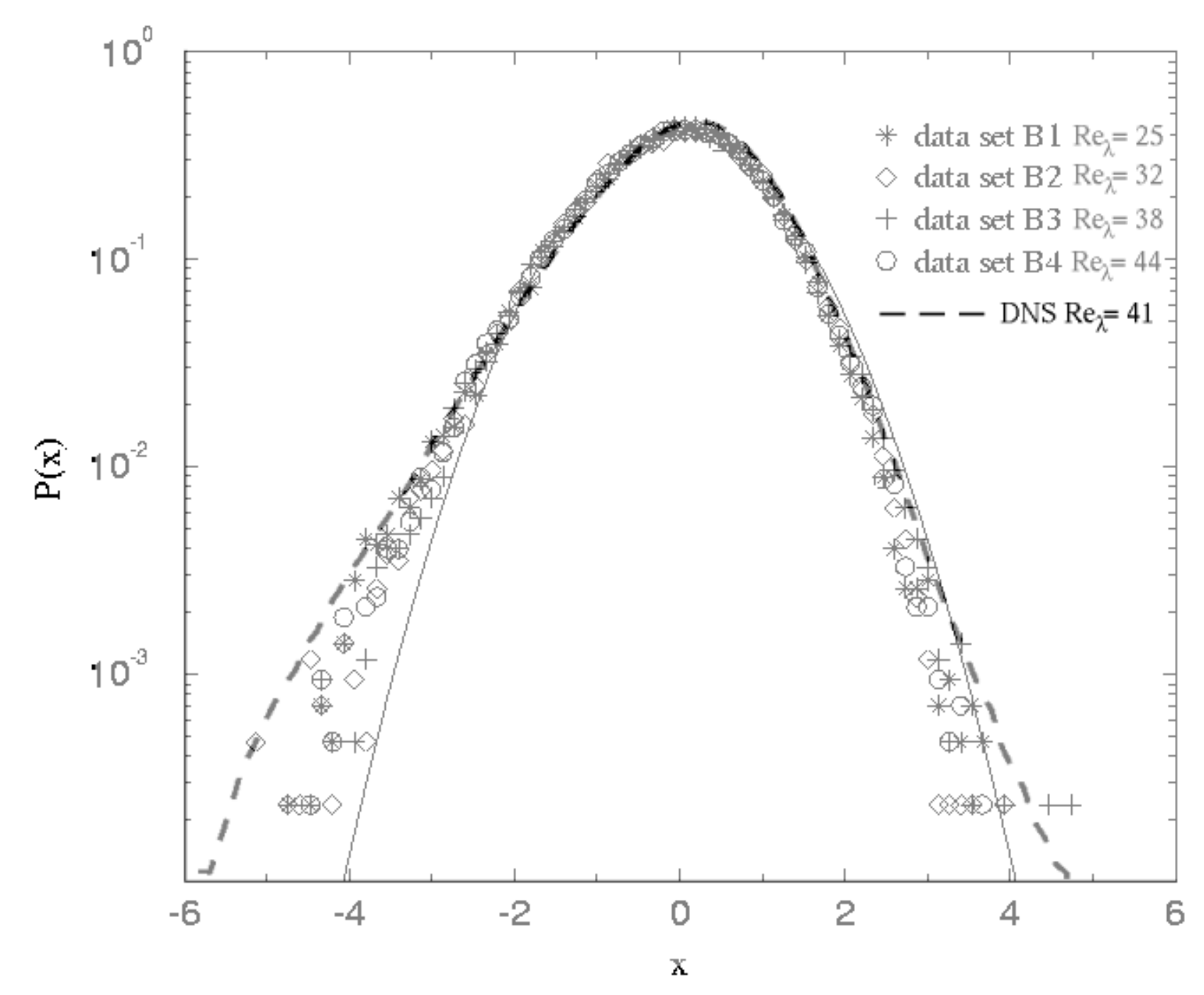}
\includegraphics[width = 80mm, angle = 0]{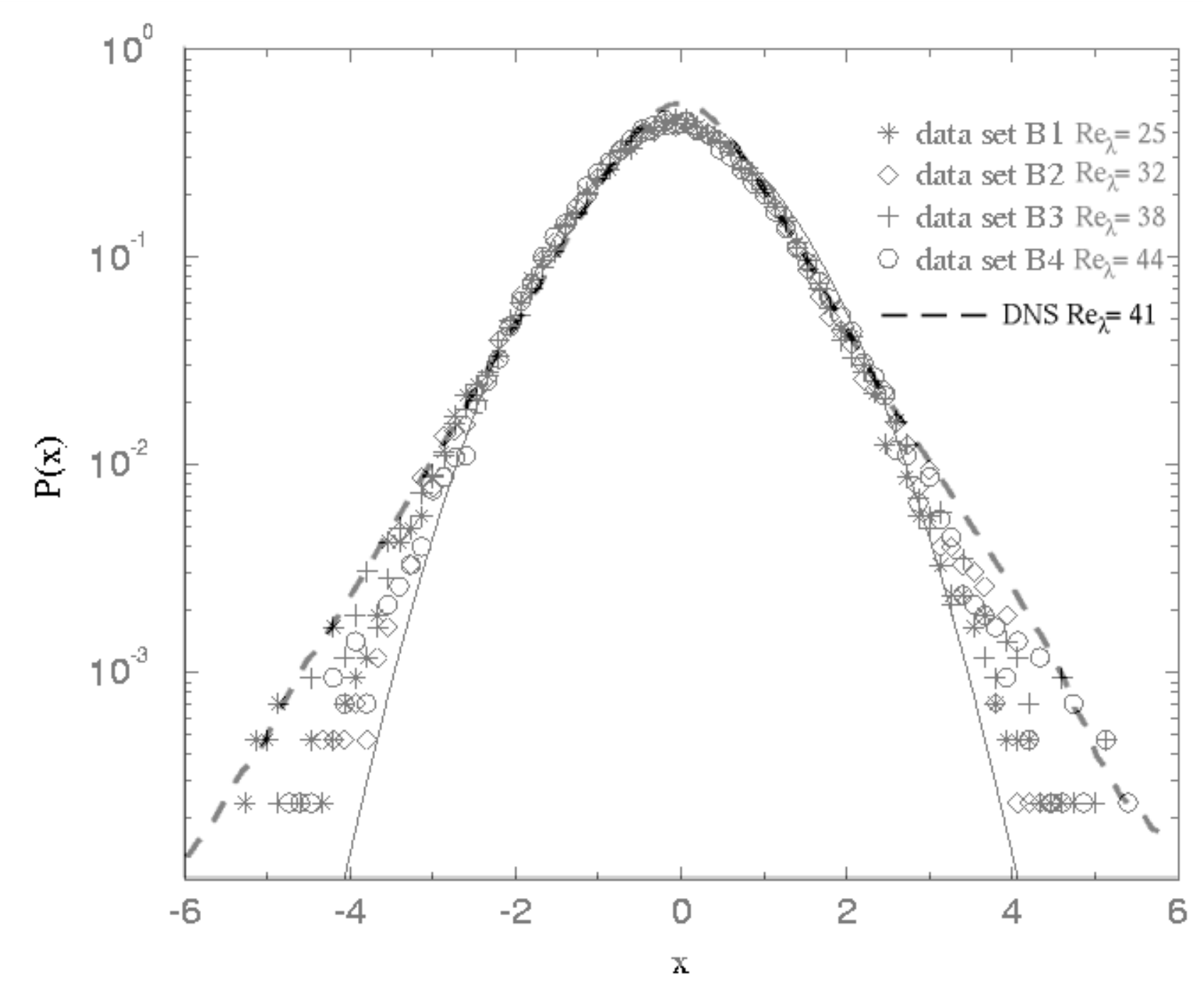} 
\caption{Probability distribution functions determined from DNS data ($Re_\lambda=41$) compared to experimental data and the Gaussian distribution. Velocity (top left); vorticity (top right); longitudinal velocity gradient (bottom left) and tranverse velocity gradient (bottom right)}\label{fig:pdfs}
\end{figure}

From the DNS data it is also possible to obtain probability density functions for comparison to the experimental data obtained. The results are shown in Figure \ref{fig:pdfs}. In all cases the distributions obtained from numerical results exhibit similar features to the distributions obtained from experimental results. 

\section{Conclusion}
The experimental measurement techniques used in this study of grid turbulence has yielded instantaneous spatial measurements of velocity, velocity gradients and vorticity. From these measurements quantities were obtained that help describe the spatial structure of an approximately homogeneous isotropic flow.

The velocity statistics, such as the decay data and the PDFs, compared favourably to previous hot-wire experimental results of grid turbulence. The values for kurtosis and skewness for the PDFs confirmed the Gaussian nature of these distributions. The velocity autocorrelation however, differed from previous experimental work. Due to the restriction of the spatial domain the influence of large scale structures present in the flow appeared to have been somewhat removed in the autocorrelations. This effect was confirmed, particularly for the case of the longitudinal autocorrelation of velocity, by examining the effect of restricting the spatial domain on velocity autocorrelations obtained from DNS data.

The PDFs of both the velocity gradients and the vorticity were found to be non-Gaussian. This result conforms with numerical results obtained previously from simulations of isotropic turbulence. Differences between experimental and numerical results for the autocorrelation functions obtained in the longitudinal direction for velocity gradient and the transverse direction for vorticity were found to be consistent with the effect found for the longitudinal autocorrelation of velocity, i.e. a result of restricting the spatial domain.

Comparing the numerical and experimental results obtained for the transverse autocorrelation function for velocity and velocity gradient indicated that the discrepancy cannot be entirely explained by the restricted spatial domain, and further that the effect of restricting the spatial domain on the autocorrelation function is unpredictable. 

The PIV technique used provided adequate spatial resolution of the turbulent scales of interest. The maximum spatial resolution achieved was approximately $0.07\lambda$, which exceeds that previously achieved in investigations of grid turbulence. This resolution has resulted in  vorticity measurements that have a maximum total error of 10\%. Temporal sequences of vorticity fields have provided visual verification of the measurements. These showed that structures of order $\lambda$ were resolved in the image plane. The challenge is now to maintain this spatial resolution while simultaneously increasing the spatial domain to obtain more accurate autocorrelation functions.

\pagebreak

\noindent{\bf References}\\
\noindent
Antonia, R.A., Phan-Thien, N., Chambers, A.J., Taylor's hypothesis and the probability density functions of temporal velocity and temperature derivatives in a turbulent flow, {\em J. Fluid Mech.} \textbf{100}, 193-208 (1980)\\
\noindent
Batchelor, G.K., Townsend, A.A., Decay of vorticity in isotropic turbulence, {\em Proc. R. Soc. Lond.} \textbf{A190}, 534-550 (1947)\\
\noindent
Batchelor, G.K., Townsend, A.A., Decay of isotropic turbulence in the initial period, {\em Proc. R. Soc. Lond.} \textbf{A193}, 539-558 (1948)\\
\noindent
Champagne, F.H., Harris,V.G., Corrsin, S., Experiments on nearly homogeneous shear flow, {\em J. Fluid Mech.} \textbf{41}, 81-139 (1970)\\
\noindent
Comte-Bellot, G., Corrsin, S., Simple Eulerian time correlation of full and narrow-band velocity signals in grid generated, ``isotropic'' turbulence, {\em J. Fluid Mech.} \textbf{48}, part 2, 273-337 (1971)\\
\noindent
Frenkiel, F.N., Klebanoff, P.S., Haung, T.T., Grid turbulence in air and water, {\em Phys. Fluids} \textbf{22}, 1606-1617 (1979)\\
\noindent
Goldstein, S., Modern developments in fluid mechanics - Volume 1, {\em Dover Publications} (1965)\\
\noindent
Hinze, J.O., Turbulence, {\em McGraw Hill} (1975)\\
\noindent
Jimenez, J., Wray, A.A., Saffman, P.G., Rogallo, R.S., The structure of intense vorticity in isotropic turbulence, {\em J. Fluid Mech.} \textbf{225}, 65-90 (1993)\\
\noindent
Kit, E., Tsinober, A., Tietal. M., Balint, J.L., Wallace, J.M., Levich, E., Vorticity measurements in turbulent grid flows, {\em Fluid Dynamics Research} \textbf{3}, 289-294 (1988)\\
\noindent
Mohamed, M.S., LaRue, J.C., The decay power law in grid-generated turbulence, {\em J. Fluid Mech.} \textbf{219}, 195-214 (1990)\\ 
\noindent
Nicolaides, D., Measurement of spatial quantities in grid turbulence using particle image velocimetry, Masters Thesis, {\em Monash University, Melbourne, Australia} (1997)\\
\noindent
O'Neill, P.L., Soria, J., The relationship between the topological structures in turbulent flow and the distribution of a passive scalar with an imposed mean gradient, submitted to {\em Fluid Dynamics Research} (2004)\\
\noindent
Snyder, W.H., Lumley, J.L., Some measurements of particle velocity autocorrelation functions in turbulent flow, {\em J. Fluid Mech.} \textbf{48}(2), 41-71 (1971)\\
\noindent
Soria, J., An adaptive cross-correlation digital PIV technique for unsteady flow investigations, {\em First Australian Conference on Laser Diagnostics in Fluid Mechanics and Combustion} 29-45 (1996a)\\
\noindent
Soria, J., An investigation of the near wake of a circular cylinder using a video-based digital cross-correlation particle image velocimetry technique. {\em Experimental Thermal and Fluid Science}, \textbf{12}(2), 221–233. http://doi.org/10.1016/0894-1777(95)00086-0 (1996b)\\
\noindent
Soria, J., Fouras, A., Accuracy of out-of-plane vorticity component measurements using in-plane velocity vector field measurements, {\em Proc. Twelfth Australasian Fluid Mechanics Conference} 387-390 (1995)\\
\noindent
Fouras, A., \& Soria, J., Accuracy of out-of-plane vorticity measurements using in-plane velocity vector field data, {\em Experiments in Fluids}, \textbf{25}, 409–430 (1998)\\
\noindent
Tennekes, H., Lumley, J.L., A first course in turbulence, {\em MIT press} (1972)\\
\noindent
Townsend, A.A., The measurement of double and triple correlation derivatives in isotropic turbulence, {\em Proc. Camb. Phil. Soc.}  \textbf{44}, 560 (1947)\\
\noindent
Townsend, A.A., The structure of turbulent shear flow, {\em Cambridge University Press} (1976)\\
\noindent
Tsinober, A., Kit, E., Dracos,T., Experimental investigation of the field of velocity gradients in turbulent flows, {\em J. Fluid Mech.} \textbf{242}, 169-192 (1992)\\
\noindent
Vincent, A., Meneguzzi, M., The spatial structure  and statistical properties of homogeneous turbulence, {\em J. Fluid Mech.} \textbf{225}, 1-20 (1991)\\
\noindent
Willert, C.E. and Gharib, M., Digital particle image velocimetry, {\em Exp Fluids} \textbf{10}, 181-193 (1991)\\
\noindent
Yeung, P.K., Pope, S.B., Lagrangian statistics from numerical simulations of isotropic turbulence, {\em J. Fluid Mech.} \textbf{207}, 531-586 (1989)\\

\end{document}